\documentclass[journal]{IEEEtran}

\usepackage{cite}
\usepackage{amsmath,graphicx}
\usepackage{times}
\usepackage[caption=false]{subfig}
\usepackage{amssymb}
\usepackage{algorithm,algorithmic}
\usepackage{epsfig}
\usepackage{multirow}
\usepackage{authblk}
\usepackage[american]{babel}
\usepackage[pagebackref=true,breaklinks=true,colorlinks,bookmarks=false]{hyperref}
\usepackage{array}
\usepackage{enumitem}
\usepackage{balance}

\newcolumntype{P}[1]{>{\centering\arraybackslash}p{#1}}
\newcolumntype{M}[1]{>{\centering\arraybackslash}m{#1}}
\DeclareMathOperator*{\argmin}{arg\,min}
\newcommand{\js}[1]{\textcolor{black}{#1}}
\newcommand{\hxy}[1]{\textcolor{black}{#1}}
\newcommand{\hxytwo}[1]{\textcolor{black}{#1}}

\begin{document}
%
\title{\hxy{Partition-Aware Adaptive Switching Neural Networks for Post-Processing in HEVC}}
%
%

\author{Weiyao~Lin, Xiaoyi~He, Xintong~Han, Dong~Liu, John~See, Junni~Zou, Hongkai~Xiong, and~Feng~Wu,~\IEEEmembership{Fellow,~IEEE}
\thanks{The basic idea of this paper appeared in our conference version~\cite{he2018enhancing}. In this
version, we extend our approach by introducing an adaptive-switching scheme, carry out detailed analysis, and present more performance results.}
\thanks{W. Lin, X. He, and H. Xiong are with the Department of Electronic Engineering,
Shanghai Jiao Tong University, China (Email: \{wylin,~515974418,~xionghongkai\}@sjtu.edu.cn).}
\thanks{X. Han is with Huya Inc., China (Email: xintong@umd.edu).}
\thanks{J. See is with the Faculty of Computing and Informatics, Multimedia
University, Malaysia (Email: johnsee@mmu.edu.my).}
\thanks{J. Zou is with the Department of Computer Science and Engineering, Shanghai Jiao Tong University, China (Email: zou-jn@cs.sjtu.edu.cn).}
\thanks{D. Liu and F. Wu are with the School of Information Science and Technology, University of Science and Technology of China (Email: \{dongeliu,~fengwu\}@ustc.edu.cn).}
}

%
%

\markboth{Prepared for IEEE Transactions on}%
{Shell \MakeLowercase{\textit{et al.}}: Bare Demo of IEEEtran.cls for IEEE Journals}
%



\maketitle

\begin{abstract}
\hxy{This paper addresses neural network based post-processing for the state-of-the-art video coding standard}, High Efficiency Video Coding (HEVC). We first propose a partition-aware Convolution Neural Network (CNN) that utilizes the partition information produced by the encoder to assist in the post-processing.
In contrast to existing CNN-based approaches, which only take the decoded frame as input,
the proposed approach considers the coding unit (CU) size information and combines it with the distorted decoded frame such that the artifacts introduced by HEVC are efficiently reduced. We further introduce an {adaptive-switching neural network} (ASN) that consists of multiple independent CNNs to adaptively handle the variations in content and distortion within compressed-video frames, providing further reduction in visual artifacts.
Additionally, an iterative training procedure is proposed to train these independent CNNs 
attentively on different local patch-wise classes. 
Experiments on benchmark sequences demonstrate the effectiveness of our partition-aware and \hxy{adaptive-switching neural networks}. The project page can be found in \href{http://min.sjtu.edu.cn/lwydemo/HEVCpostprocessing.html}{http://min.sjtu.edu.cn/lwydemo/HEVCpostprocessing.html}.
\end{abstract}

\begin{IEEEkeywords}
High Efficiency Video Coding, Convolutional neural network, Post-processing
\end{IEEEkeywords}

%
\IEEEpeerreviewmaketitle

\section{Introduction}
\label{sec:intro}

Recently, the fast development of video capture and display devices has brought a dramatic demand for high definition (HD) contents. Thus, High Efficiency Video Coding (HEVC) \cite{hevc} standard is developed by the Joint Collaborative Team on Video Coding (JCT-VC). HEVC provides higher compression performance compared to the previous standard H.264/AVC by 50\% of bitrate saving on average at a similar perceptual image quality \cite{Comparison_of_the_coding}. However, HEVC videos still contain compression artifacts, such as blocking artifacts, ringing effects, blurring, \textit{etc.}, making it of great importance to study on reducing the visual artifacts of the decoded videos.

Over the past years, there has been a lot of work aiming at reducing the visual artifacts of decoded images \cite{liew2004blocking, jancsary2012loss,jung2012image,wang2013adaptive,chang2014reducing} and videos \cite{han2015high,zhang2016structure,li2016lagrangian}. Inspired by the great success achieved by deep learning techniques in computer vision and image processing tasks \cite{krizhevsky2012imagenet,girshick2014rich,kim2016accurate}, many convolutional neural network based approaches \cite{arcnn,wang2016d3,ifcnn,vrcnn,DCAD,qecnn,DRN2019,MSDD} have been proposed to mitigate the visual artifacts in decoded images and videos. 

\begin{figure}[t]
 \centering
\subfloat[]{\includegraphics[width=0.48\linewidth]{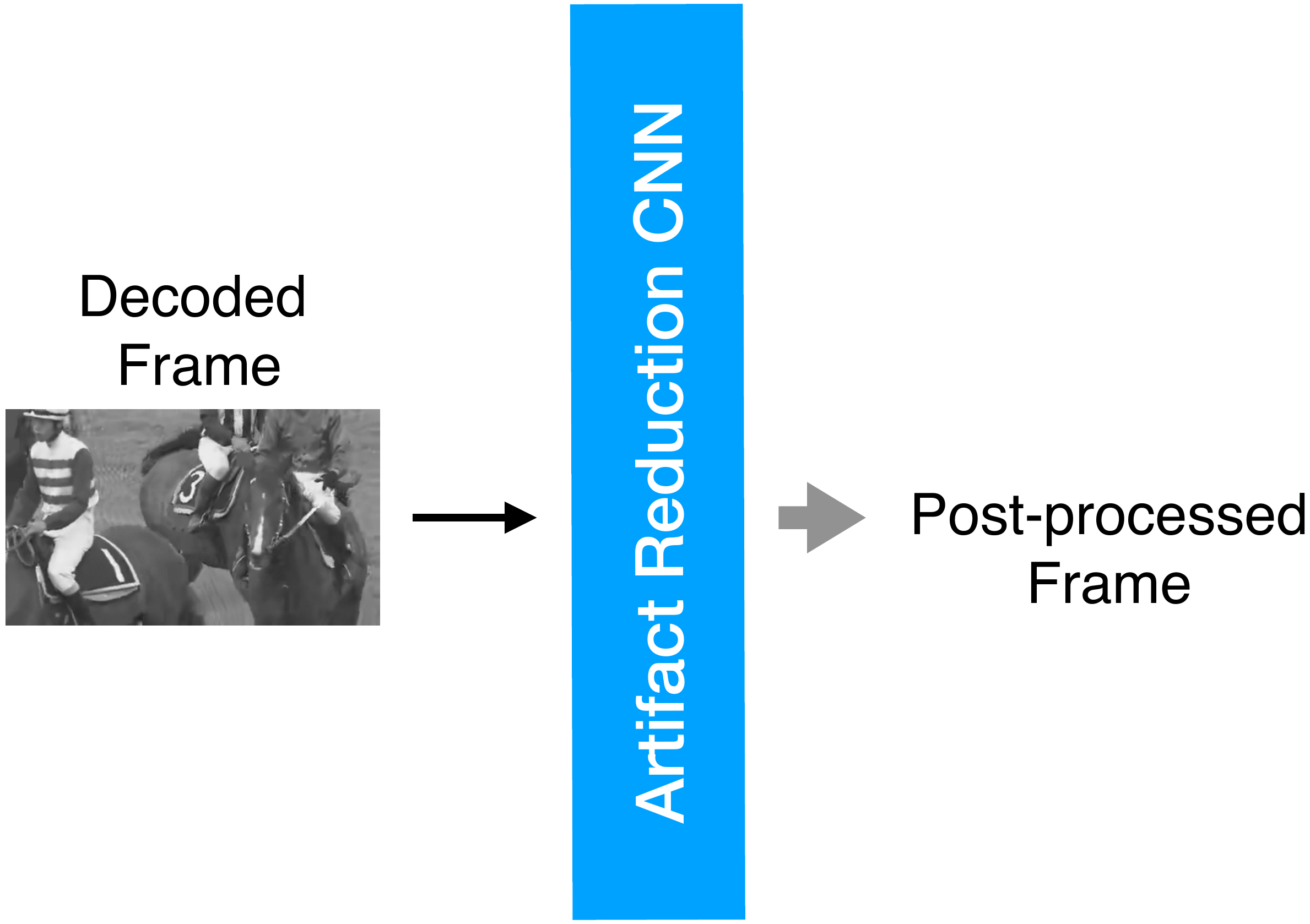}\label{fig:overview:a}}
\hfill
 \subfloat[]{\includegraphics[width=0.48\linewidth]{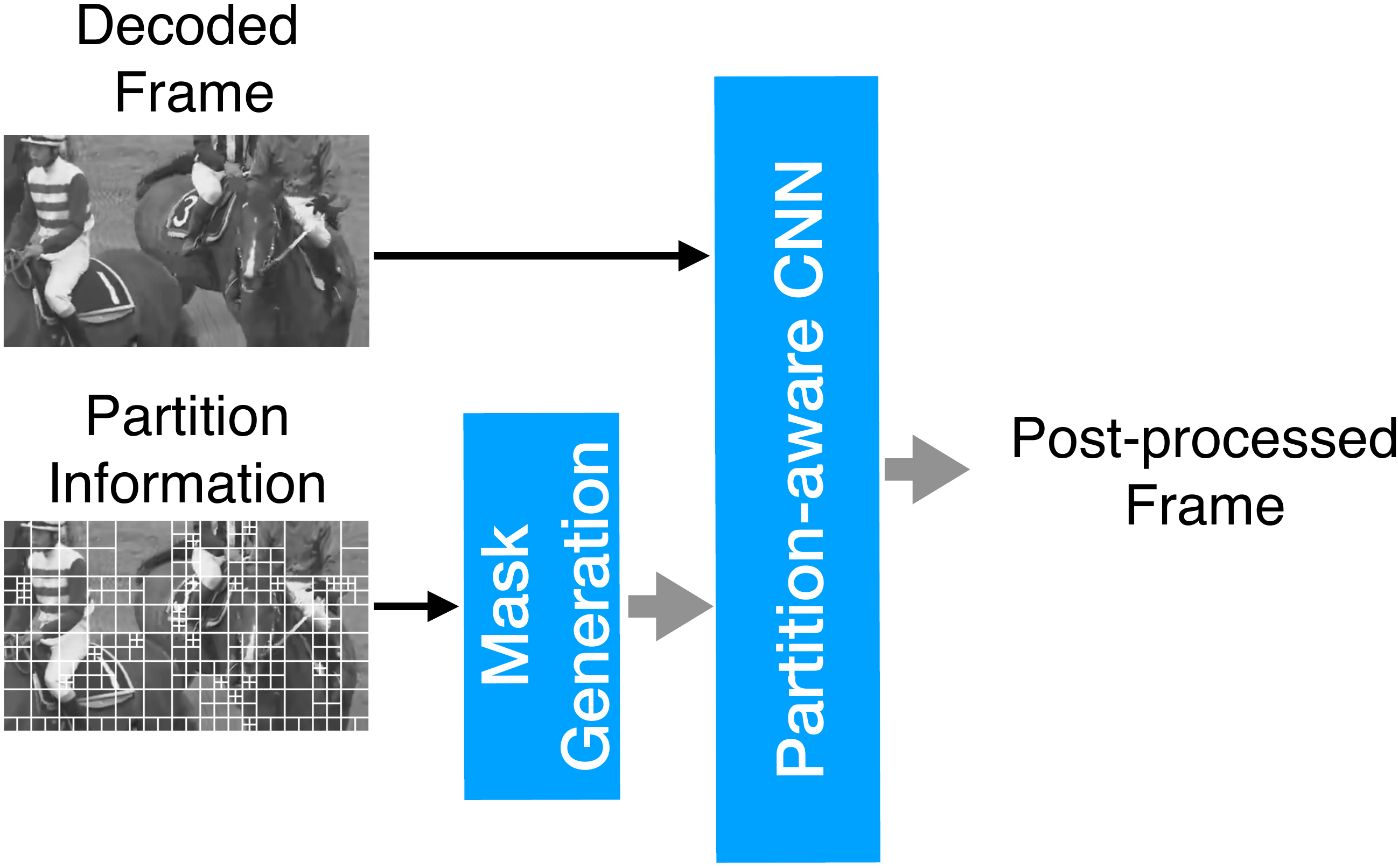}\label{fig:overview:b}}
\hfill
\subfloat[]{\includegraphics[width=0.4\linewidth]{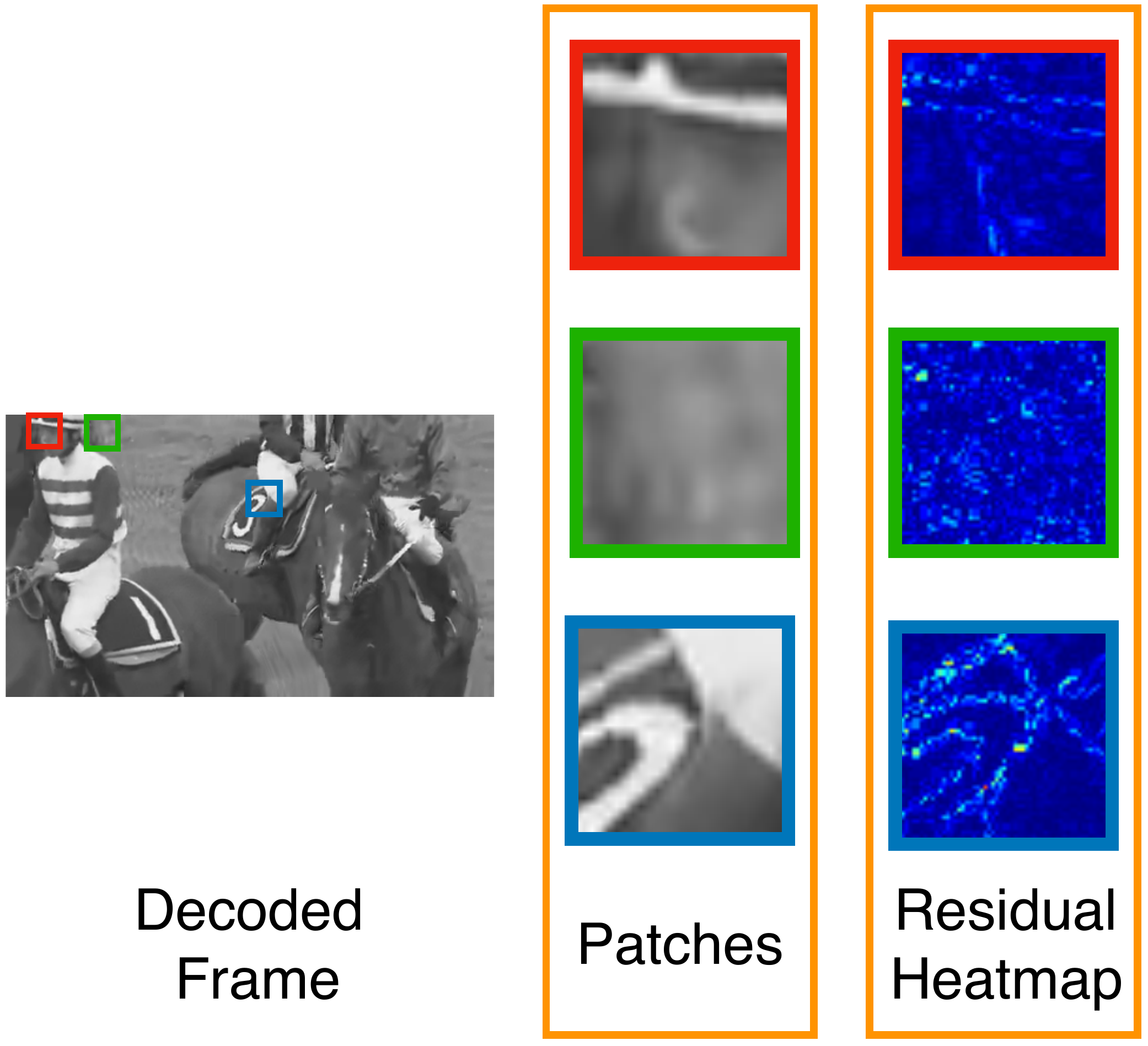}\label{fig:overview:c}}
\hspace{0.05\linewidth}
\subfloat[]{\includegraphics[width=0.4\linewidth]{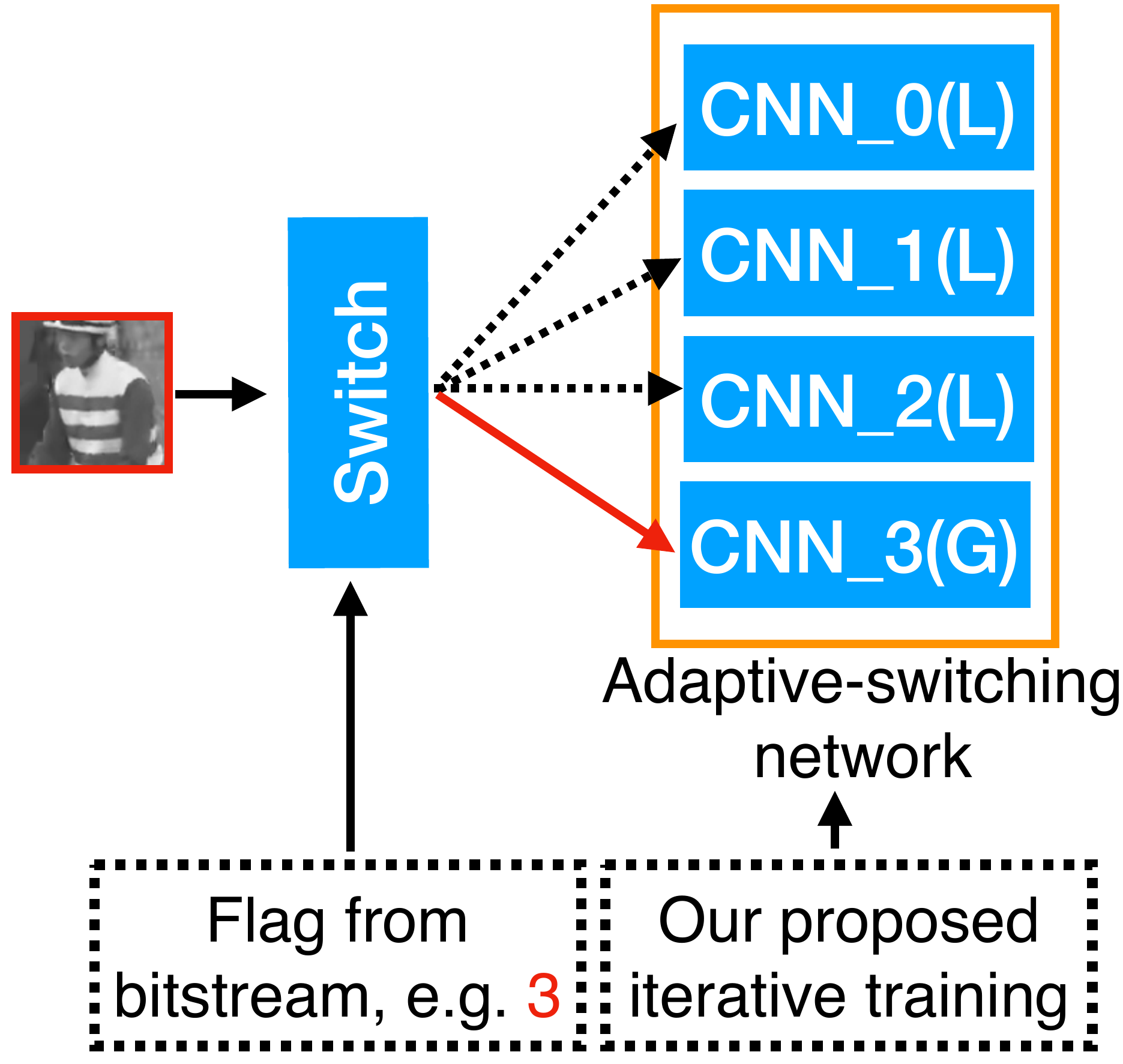}\label{fig:overview:d}}
\hfill
 \caption{(a) Traditional single input methods; (b) Our partition-aware CNN; (c) Heatmap of residual; (d) Our \hxy{adaptive-switching neural network}.}
\end{figure}

 \begin{figure*}[ht]
\centering
\subfloat[]{\includegraphics[height=5cm]{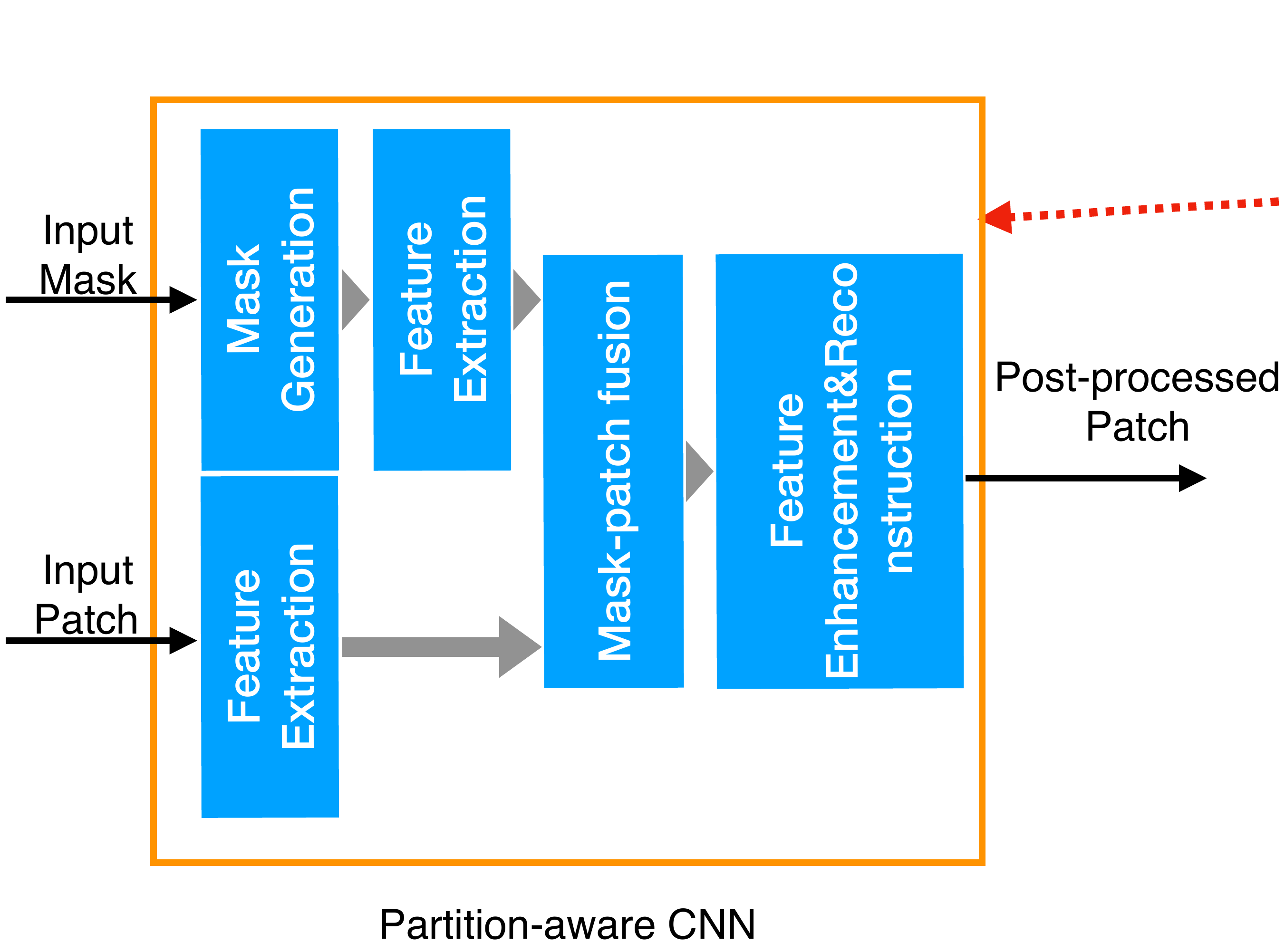}\label{fig:framework_a}}
\subfloat[]{\includegraphics[height=5cm]{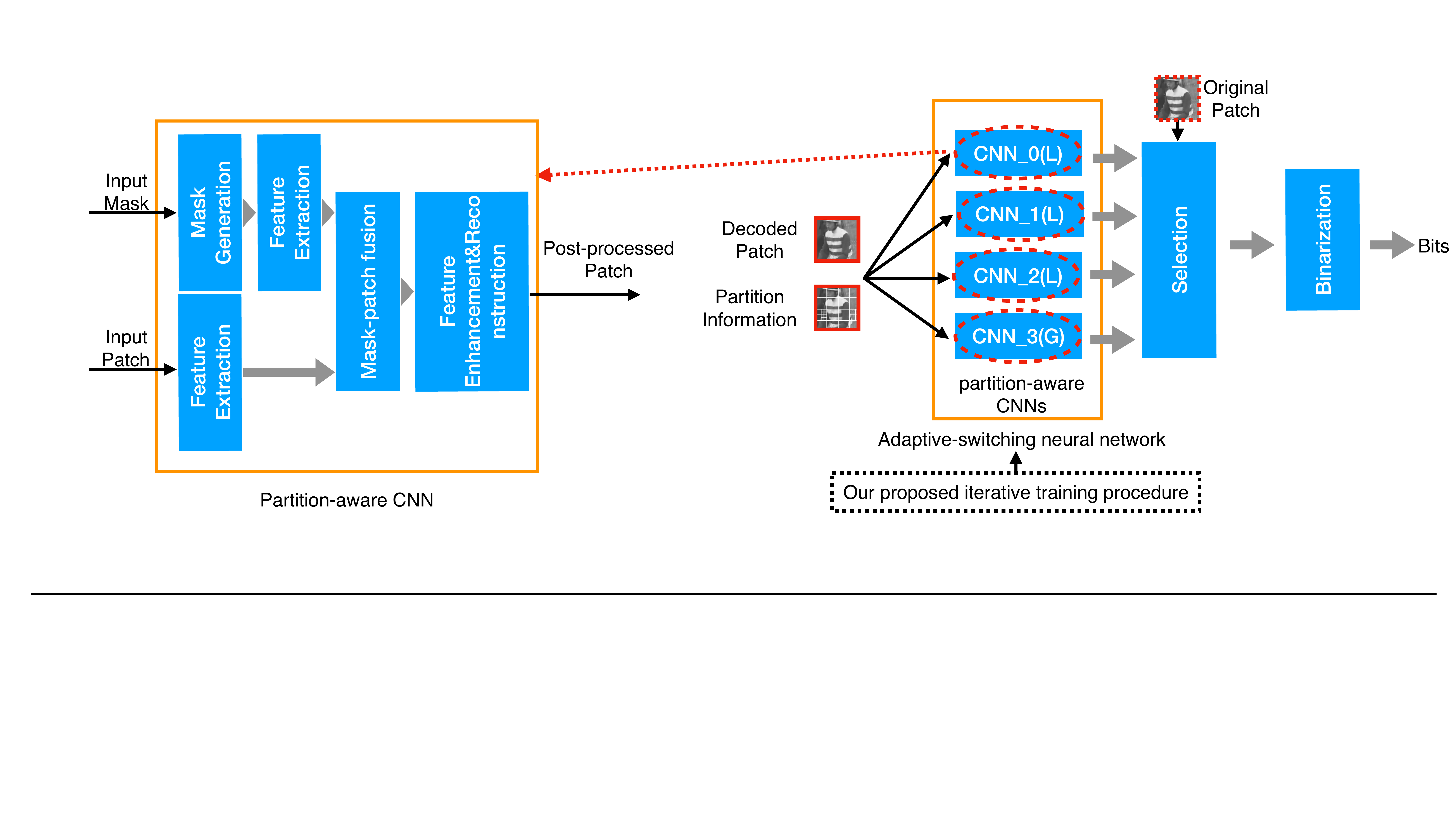}\label{fig:framework_b}}
\hfill
\caption{The framework of our proposed method. (a) Partition-aware CNN; (b) Adaptive-switching scheme.}
\label{fig:framework}
\end{figure*}
As for reducing the visual artifacts of HEVC compressed videos, \cite{ifcnn, vrcnn, DCAD, qecnn, DRN2019, MSDD} introduced a set of CNN-based filters which post-process HEVC compressed frames and obtain improved visual quality by suppressing artifacts. However, most existing methods~\cite{ifcnn, vrcnn, DCAD, DRN2019, MSDD} only take the decoded patches as the input to the CNN and output the post-processed patches directly as shown in Fig.~\ref{fig:overview:a}, while no other prior information is taken into consideration explicitly. In this paper, we posit that partition information (\textit{e.g.}, 16$\times$16, 8$\times$8~\cite{hevc}, shown in Fig~\ref{fig:overview:b}) could be used to effectively guide the post-processing performed by CNN. In practice, since the partition information is introduced by the block-wise processing and quantization of HEVC, it is related to the content of the frame and indicates the source of visual compression artifacts.

Based on this intuition, we propose a novel approach (shown in Fig.~\ref{fig:overview:b}) that first derives a carefully designed mask from a frame's partition information, and then uses it to guide the post-processing of the decoded frame through a partition-aware CNN. Note that the partition information (e.g., 16$\times$16, 8$\times$8) is generated already by encoder~\cite{shen2013effective}. As a result, the visual artifacts of HEVC-compressed videos can be more effectively reduced under the same bit rate.

Moreover, the recent study~\cite{qecnn} attempts to use two CNNs with different architectures to handle degradation variation in HEVC intra and inter coding. Though it has the specific architectures for HEVC intra- and inter-coding frames, the degradation variation within decoded frames (\textit{e.g.}, among patches) is neglected. Fig.~\ref{fig:overview:c} shows examples of patches and their corresponding residual heatmaps from one decoded frame. The residual is the absolute difference between the decoded and original frames (only Y-channel is shown). We can see obvious difference among these three patches' residual heatmaps. This indicates different patches may need to be treated differently for post-processing. To this end, we introduce an adaptive-switching scheme to handle degradation variation within decoded frames, which leads to further performance improvement. Its framework in decoder side is shown in Fig. \ref{fig:overview:d}. Patches are relayed to independent CNNs in an \hxy{adaptive-switching neural network} (ASN) based on flags read from the bitstream. The designed ASN consists of multiple CNNs, where we let \textit{CNN\_i(L)} denote the local CNN with index $i$ ($i \in [0,2]$) and \textit{CNN\_3(G)} denote the global CNN. Note that all local CNNs are trained by our proposed iterative training method to enable each of them focus more on the specific class of local patches and the global CNN is trained on the whole training dataset (see Section \ref{sec:iterative_training} for more details).

 \begin{figure*}[t]
    \centering
  \subfloat[Concatenation-based late fusion]{\includegraphics[width=0.32\linewidth]{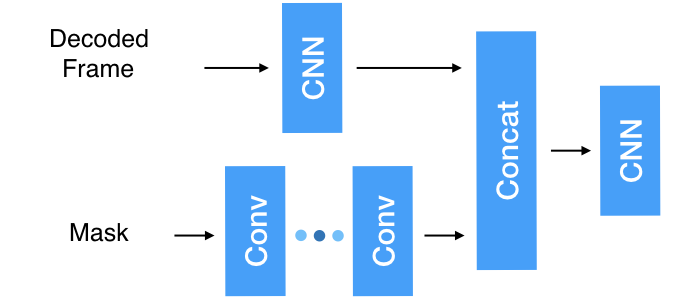}\label{fig:4c}}
  \hfill
  \subfloat[Addition-based fusion.]{\includegraphics[width=0.32\linewidth]{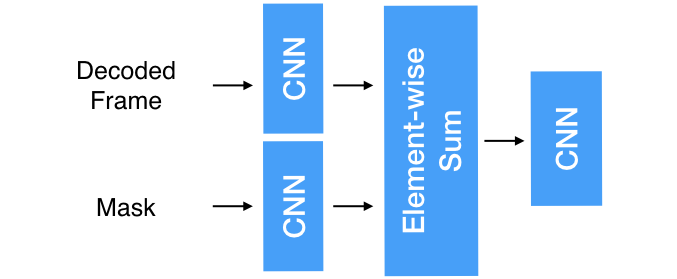}\label{fig:4a}}
  \hfill
  \subfloat[Concatenation-based early fusion]{\includegraphics[width=0.32\linewidth]{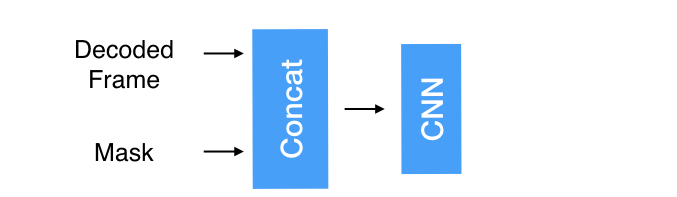}\label{fig:4b}}
  \hfill
  \caption{Different fusion methods to combine the decoded frame and mask.}
  \label{fig:fusin_str}
\end{figure*}

 \begin{figure}[t]
\centering
\subfloat[Original frame with partition information]{\includegraphics[width=0.32\linewidth]{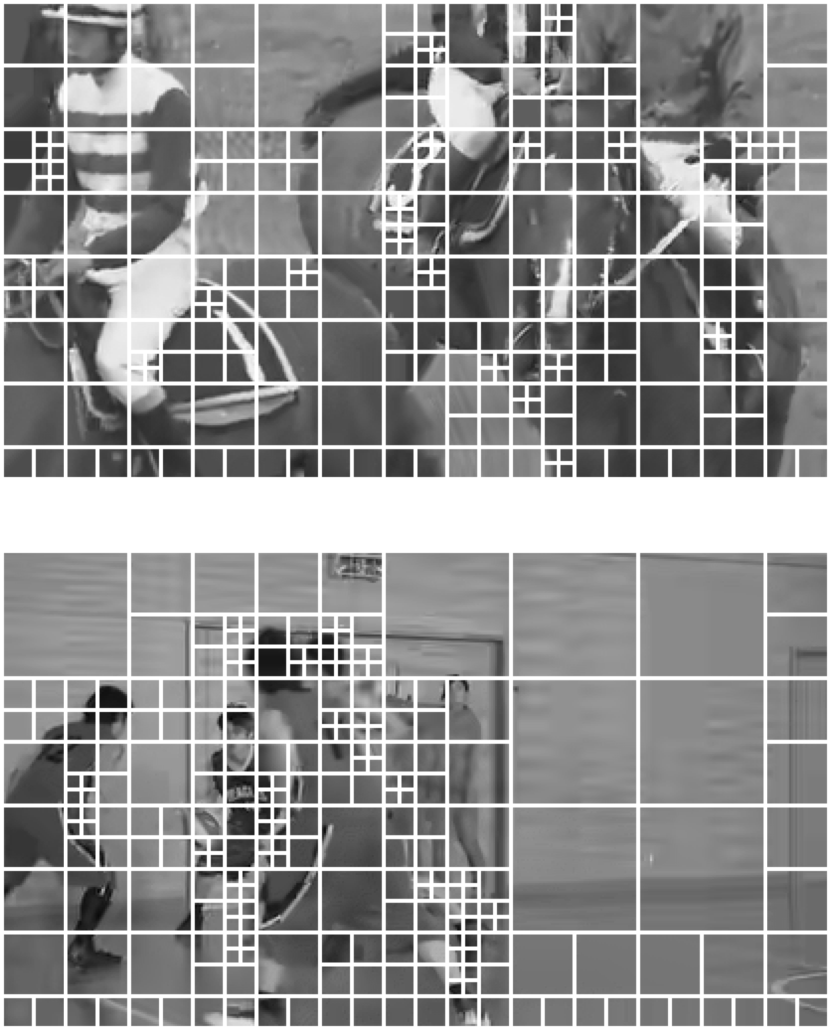}\label{fig:3a}}
\hfill
\subfloat[Local Mean-based \newline mask]{\includegraphics[width=0.32\linewidth]{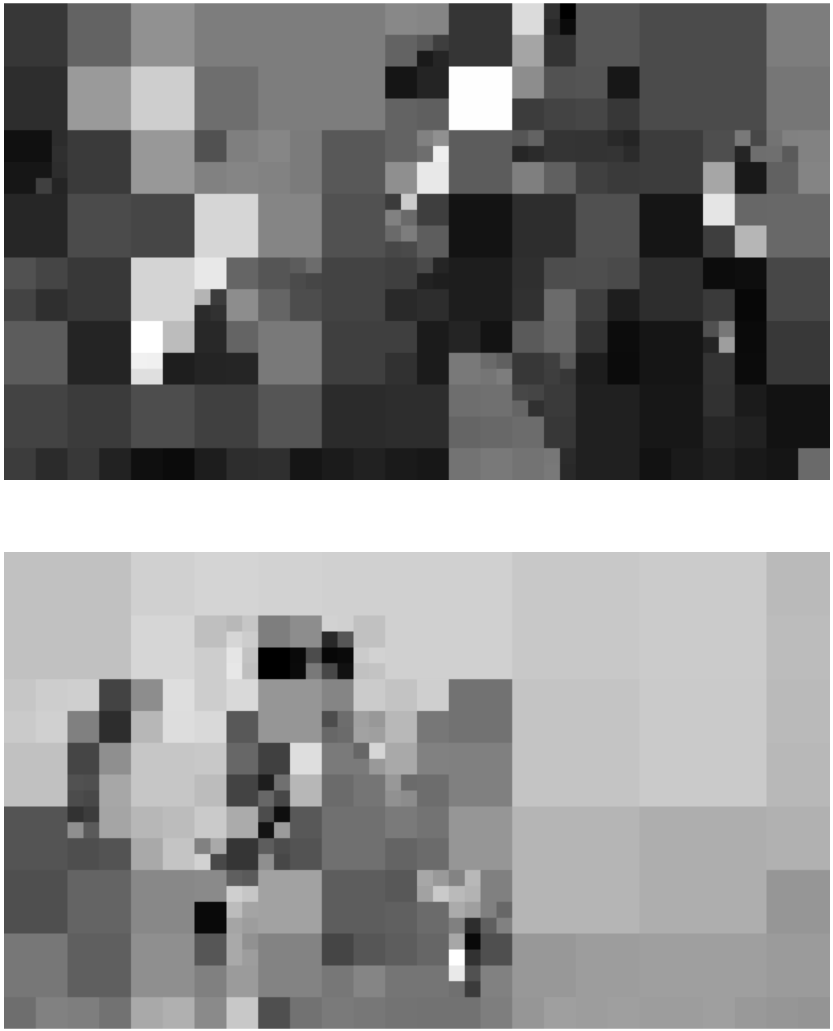}\label{fig:3b}}
\hfill
\subfloat[Boundary-based \newline mask]{\includegraphics[width=0.32\linewidth]{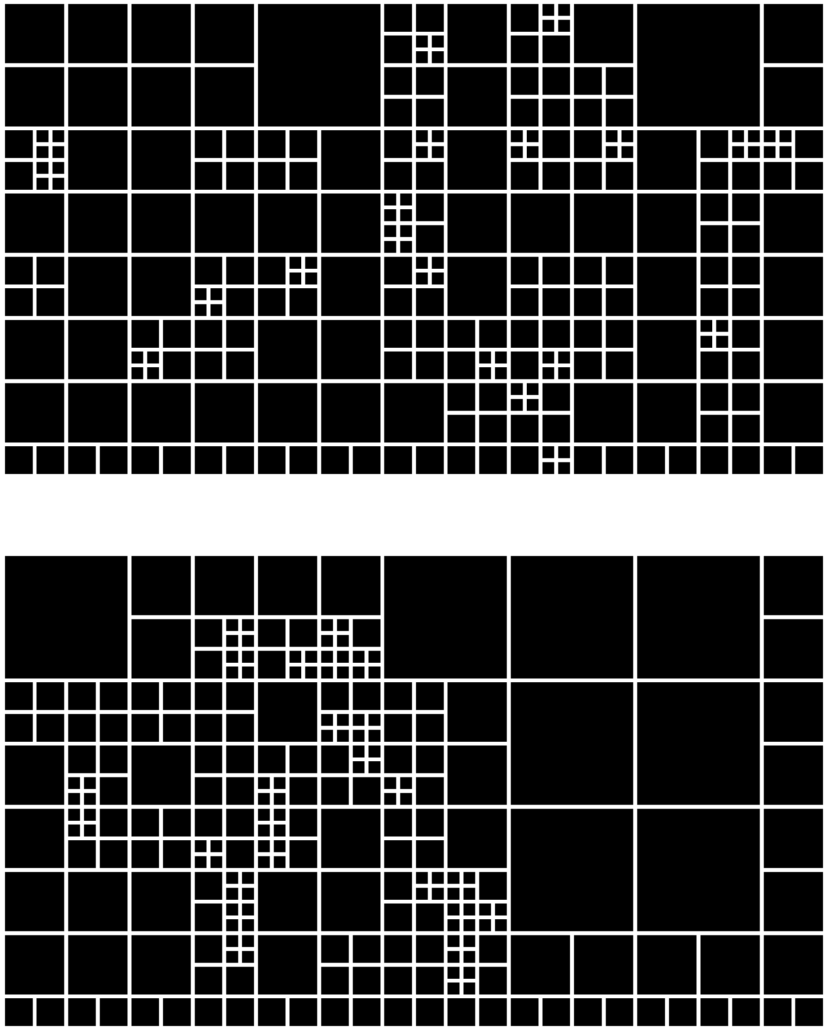}\label{fig:3c}}%
\hfill
\caption{Two examples of boundary-based mask and local mean-based mask.}
\label{fig:mask}
\end{figure}

In summary, our contributions are three-fold:
\begin{enumerate}[noitemsep]
\item We develop a novel framework that utilizes the partition information to guide the CNN-based post-processing in HEVC, where a mask derived from a decoded frame's partition information is fused with this decoded frame through a partition-aware CNN to accomplish post-processing. Besides, under this framework, we systematically investigate different mask generation and mask-frame fusion methods and find the best strategies. We also demonstrate that our approach is general and can be integrated into the existing HEVC compressed-video post-processing methods to further improve their performances.

\item Our adaptive-switching scheme utilizes multiple CNNs (one global and a set of local CNNs) to handle degradation variation of local patches within decoded frames and further reduce the visual artifacts. Moreover, all local CNNs are trained by our carefully designed iterative training strategy, so that each of them concentrates on specific class of local patches. We conduct experiments by applying this scheme and training method to demonstrate their effectiveness with different CNN architectures.

\item We establish a large-scale dataset containing 202,251 training samples for encouraging training compressed-video post-processing models. This dataset will be made publicly available to facilitate further research.


\end{enumerate}

\textbf{The frameworks of our work.} The frameworks of the proposed partition-aware CNN and adaptive-switching scheme are shown in Fig~\ref{fig:framework}. Our partition-aware CNN, which is detailed in Section~\ref{sec:detail}, is shown in Fig~\ref{fig:framework_a}. For each patch in a decoded frame, we obtain its corresponding mask generated by the patch's partition information, and fed this information together with the patch into a partition-aware CNN. Inside this CNN, the features of the mask and decoded patch are first extracted through two individual streams and then fused into one. The rest layers of the partition-aware CNN perform the feature enhancement, mapping, reconstruction, and output the post-processed patch with reduced artifacts.

As for our adaptive-switching scheme (detailed in Section~\ref{sec:detail2}) shown in~\ref{fig:framework_b}, each patch is post-processed by a bank of trained CNNs in the encoder side. These CNNs consist of three local CNNs (\textit{CNN\_0(L)}, \textit{CNN\_1(L)}, \textit{CNN\_2(L)}) and one global CNN (\textit{CNN\_3(G)}) . Then the CNN is chosen such that the difference (measured by PSNR~\cite{hevc}) between the post-processed patch and its original patch is smallest across all CNNs. This amounts to greedily choosing the CNN that generated the most similar one to original frame patch in terms of PSNR among all CNNs. The indices of chosen CNNs are directly written into bitstream after binarization.

The rest of this paper is organized as follows. Section II reviews related works. Sections III describes the proposed partition-aware CNN in detail. In Section IV, we describe the details of our proposed adaptive-switching scheme. Section V shows the experimental settings and results of our proposed partition-aware CNN and adaptive-switching scheme. Section VI concludes the paper.

\section{Related work}

Over the past decades, a lot of works aiming to reduce visual artifacts have been proposed. Liew \textit{et al.} \cite{liew2004blocking} proposed a wavelet-based deblocking algorithm to reduce artifacts of compressed images. Jancsary \textit{et al.} \cite{jancsary2012loss} developed a densely-connected tractable conditional random field to achieve JPEG images deblocking. Non-local means filter was applied to remove quantization noise on blocks by Wang \textit{et al.} in \cite{wang2013adaptive}.

Recently, due to the impressive achievements of deep neural networks in computer vision and image processing tasks, many deep learning based methods are proposed to further reduce the visual artifacts of decoded images \cite{arcnn,wang2016d3,ifcnn,vrcnn,qecnn}. More specifically, Dong \textit{et al.} \cite{arcnn} developed an Artifacts Reduction CNN (ARCNN) built upon \cite{srcnn}, which successfully reduces the JPEG compression artifacts. The proposed network contains four convolutional layers and takes the JPEG compressed image as input and outputs artifact-reduced decoded image directly. According to \cite{arcnn}, ARCNN shows superior performance compared with the state-of-the-art conventional methods. This inspires researches to focus on deep learning based methods for reducing visual artifacts. Following the similar line, \cite{wang2016d3} designed a Deep Dual-Domain network that takes the prior knowledge of the JPEG into consideration to remove artifacts of JPEG compressed images.

Similar to images, there are also a lot of works focusing on deep learning based artifact reduction for decoded videos \cite{ifcnn, vrcnn, qecnn}. Park and Kim \cite{ifcnn} proposed an In-loop Filter CNN (IFCNN) based on SRCNN \cite{srcnn} to replace the Sample Adaptive Offset (SAO) filter in HEVC. There are two in-loop filters in HEVC: deblocking filter and SAO filter. These in-loop filters are able to improve visual quality and decrease bit-rate of compressed videos and therefore improve coding efficiency. IFCNN outperforms the HEVC reference mode (HM) with average 1.9\% - 2.8\% bit-rate reduction for Low Delay configuration, and average 1.6\% - 2.6\% bit-rate reduction for Random Access configuration. Slightly different from \cite{ifcnn}, Dai \textit{et al.} \cite{vrcnn} designed a VRCNN based on ARCNN \cite{arcnn} as a post-processing technique to further reduce the visual artifacts of decoded frames in HEVC.

VRCNN can improve the visual quality of decoded frames without any increase in bit-rate and therefore also can improve the coding efficiency of HEVC in average 4.6\% bit-rate reduction for intra coding. To further improve coding efficiency, Wang \textit{et al.} \cite{DCAD} proposed a deeper CNN called DCAD that has 10 convolutional layers. The same model trained by intra-coding frames of HEVC are applied for both intra and inter coding during the test stage in \cite{DCAD}.

However, since these works do not consider any prior information and only take decoded frame as input, they have limitations in handling complex degradation introduced by HEVC. Recently, Yang \textit{et al.} \cite{qecnn} proposed QECNN that has specific architectures for HEVC intra- and inter- coding frames. This amounts to process I and P/B frames in HEVC with two separate models. But degradation variation within decoded frames is neglected and the potential of multiple CNNs can be further explored.

In contrast to these existing methods, our proposed partition-aware CNN takes the partition information into consideration and a mask derived from it is fed into CNN. Besides, we propose an adaptive-switching scheme considering content and degradation variation within compressed-video frames, where patches in decoder side are relayed to independent CNNs in a greedy fashion. Compared with the previous methods, the degradation introduced by HEVC can be reduced more efficiently using our partition-aware CNN. Furthermore, compared with methods based on a single CNN, artifacts can be further reduced with our adaptive-switching scheme.

\begin{figure*}[!t]
 \centering
 \includegraphics[width=0.95\linewidth]{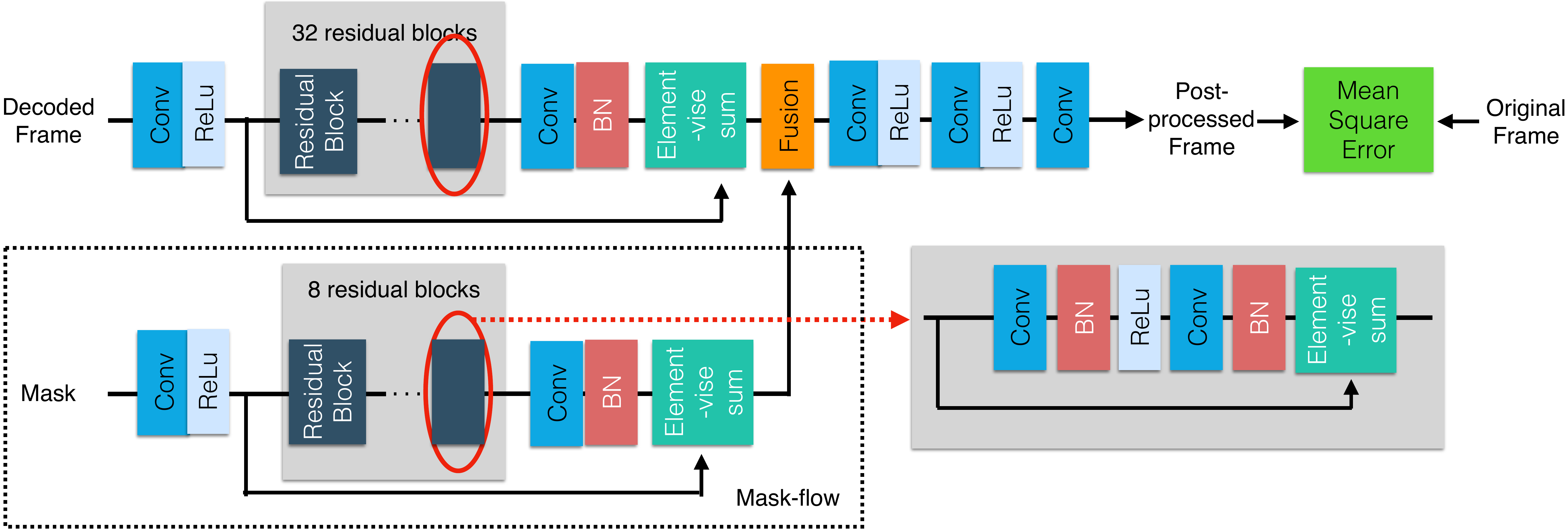}
 \caption{Our partition-aware CNN.}
 \label{fig:cnn_detail}
\end{figure*}

\section{The partition-aware network}
\label{sec:detail}
This section will first discuss the key components of our partition-aware CNN -- mask generation \& mask-patch fusion strategies, and then give details of partition-aware CNN with the specific mask generation \& mask-patch fusion strategies.

\subsection{Mask generation and mask-frame fusion strategies}
  \label{sec:mask_and_fusion}
 Since the block-wise transform and quantization are performed in HEVC during encoding, the quality degradation of compressed frames is highly related to the coding unit (CU) partition \cite{hevc}. Thus, the partition information contains useful clues for eliminating the artifacts presented during encoding. Considering this, we design a mask based on the partition information of these CUs to guide the post-processing process.

\textbf{Mask generation.} We introduce 2 strategies to generate masks from an HEVC-encoded frame's partition information:

 \begin{itemize}
  \item Local mean-based mask (MM). We fill each partition block in a frame with the mean value of all decoded pixels inside this partition. An example of a generated mean-based mask is shown in Fig. \ref{fig:3b}. As we can see that the different partition blocks are properly displayed in the mask. In this way, when we fuse it with the decoded frame during the post-processing process, it can effectively distinguish different partition modes and reduce the compression artifacts more effectively.
  \item Boundary-based mask (BM). We also introduce a boundary-based mask generation strategy. In this mask, the boundary pixels between partitions are filled with value 1 and the rest non-boundary pixels are filled with 0 (cf. Fig. \ref{fig:3c}). The width of the boundary is set to 2.
\end{itemize}

 \textbf{Mask-frame fusion strategies.} The mask is fed into CNN and integrated with its corresponding decoded frame to get the fused feature maps. We also introduce 3 strategies to fuse the information of a decoded frame and its corresponding mask:

\begin{itemize}
  \item Concatenate-based late fusion (CLF). We extract the features of mask only using three convolutional layers and integrate it into the network as shown in Fig. \ref{fig:4c}.
  \item Add-based fusion (AF). As shown in Fig. \ref{fig:4a}, we first extract the feature maps of the mask using CNN and then combine it with the feature maps of the input frame using element-wise add layer.
  \item Concatenate-based early fusion (CEF). We concatenate the mask and frame as the input to the CNN. Then the two-channel input is fed to CNN directly (cf. Fig. \ref{fig:4b}).

\end{itemize}


\subsection{Details of partition-aware convolutional neural network}
\label{sec:base_network}
In Fig. \ref{fig:cnn_detail}, we illustrate the architecture of the proposed partition-aware convolutional neural network that integrates partition information with add-based fusion strategy to reduce the visual artifacts of compressed frames. Note that the framework of our proposed partition-aware CNN is general. Besides the CNN structure in Fig. \ref{fig:cnn_detail}, it can also be integrated with other CNN structures for performing post-processing.

According to \ref{fig:cnn_detail}, the CNN contains two streams in the feature extracting stage so as to extract features for the decoded frame and its corresponding mask, respectively. Each residual block \cite{block,srgan} in the feature extracting stage has two convolutional layers with 3$\times$3 kernels and 64 feature maps, followed by batch-normalization \cite{bn} layers and ReLU activation functions. Then, the feature maps of the mask and decoded frame are fused by the add-based fusion strategy and are fed to the rest three convolutional layers. These three layers are utilized for feature enhancement, mapping, reconstruction (as described by \cite{arcnn}), and finally outputting the post-processed frame with reduced artifacts. Moreover, when training the network, the Mean Squared Error between the original raw frame and the CNN output is used as the loss function.

Compared with the existing compressed video post-processing methods \cite{vrcnn,qecnn}, our network has two differences: (1) We introduce two stream inputs to include both the decoded frame and the partition information. (2) We use residual architecture to perform the feature extraction. The deep residual stream can capture the feature of input in a more distinctive and stable way.

\section{Adaptive-switching Scheme}
\label{sec:detail2}

\js{To increase the diversity of the local representations, we propose an iterative training procedure to obtain local CNNs in the form of an \hxy{adaptive-switching neural network}.}
\js{Besides, this scheme also aids in the artifact reduction of local patches of different classes.}

\subsection{Iterative training}
\label{sec:iterative_training}
\begin{figure}[!t]
 \centering
 \includegraphics[width=\linewidth]{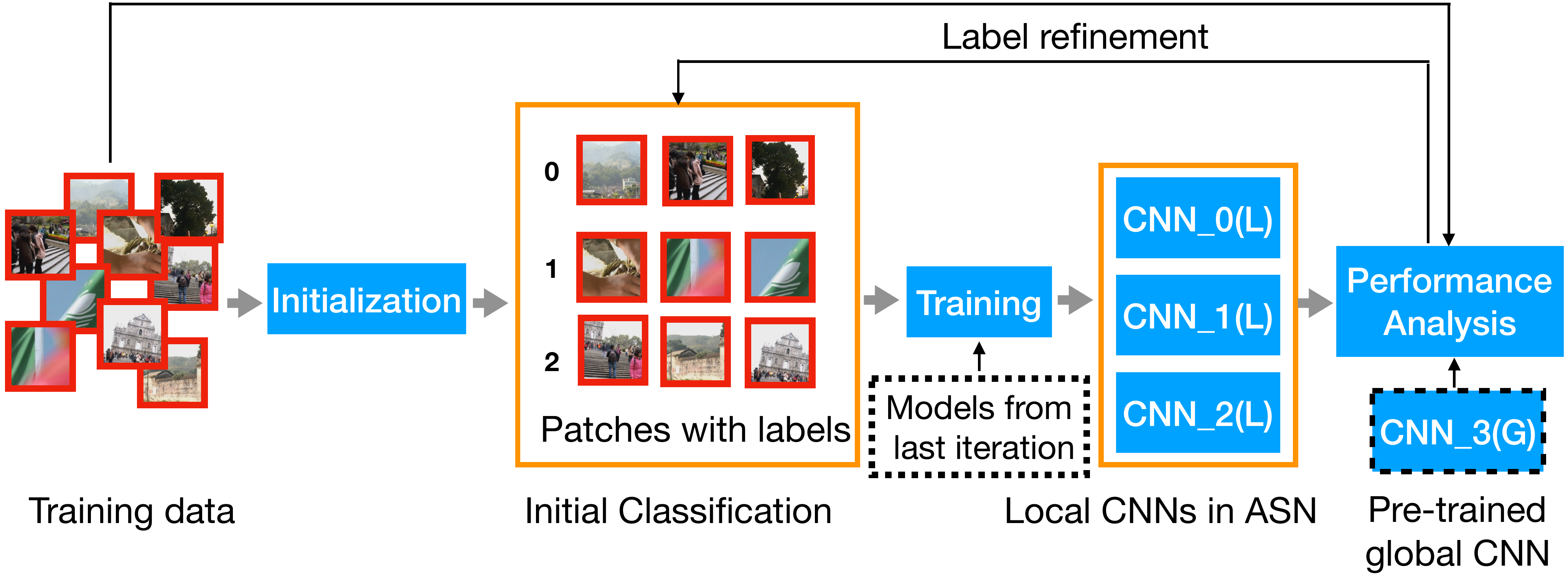}
 \caption{The framework of our iterative training.
 }
 \label{fig:iterative_training}
 \end{figure}

The framework of this iterative procedure is shown in Fig. \ref{fig:iterative_training}. At the first step, the labels of all training patches are initialized by a specific initialization method
with the number of classes set to three. \js{Their corresponding pre-trained models at this initial step are also provided.} \js{For} each iteration, each individual local CNN is fine-tuned from the model obtained from the \js{previous} iteration based on its \js{corresponding patch class.} We obtain the final local CNNs when {the training converges after} a number of epochs.
In the performance analysis \js{step}, the trained local CNNs and a pre-trained global CNN are used to generate new labels for all training patches. The pre-trained global CNN is fixed and used during performance analysis only. After label refinement, each training patch in the training set is assigned a new label. This procedure is repeated until the performance of \hxy{adaptive-switching neural network} converges \js{(with little further change)} or the maximum \js{number of} iterations is reached.

We now discuss \js{in detail} the key components \js{of our proposed scheme} - initialization methods \& the iterative process.

\subsubsection{Initialization methods} 
\begin{figure}[!t]
 \centering
 \includegraphics[width=0.8\linewidth]{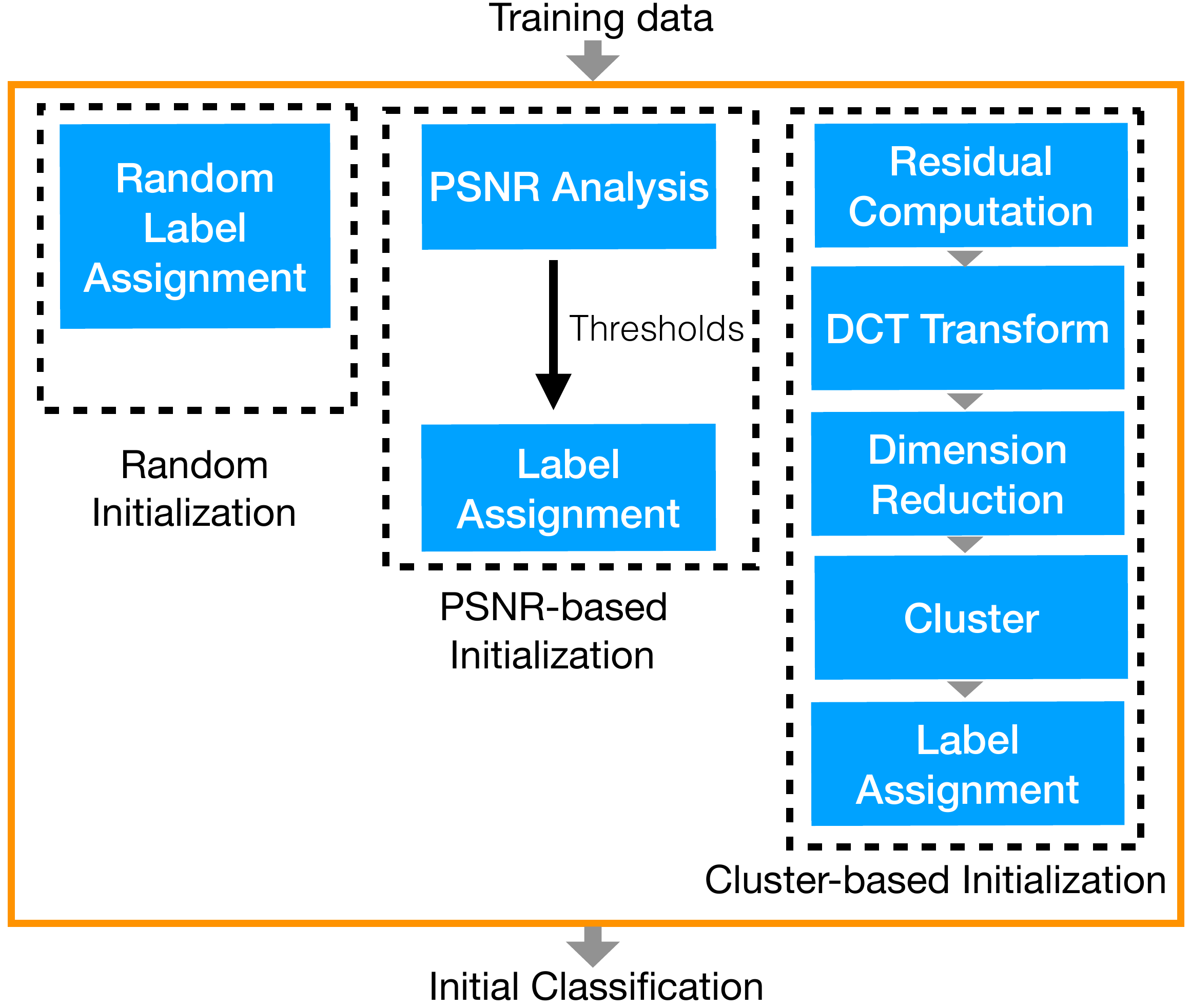}
 \caption{Different initialization methods.}
 \label{fig:ini_methods}
\end{figure}

Since we do not have the ground truth \js{label}  
\js{for each patch,} selecting a proper initialization method is key to \js{effective} iterative training. We introduce three initialization methods (as illustrated in Fig.~\ref{fig:ini_methods}): Random initialization, PSNR-based initialization method, and Cluster-based initialization method.

\begin{itemize}
\item \textbf{Random Initialization. 
} 
\js{In this method, the}
labels for all training patches are initialized 
at random.
\item \textbf{PSNR-based initialization method. 
} 
\js{Using} the PSNR of all training patches, we pick two thresholds that split the training patches into three \js{quasi-equal} parts. Then, \js{we consider} each part as one \js{patch} class of training data. Note that this initialization method is essentially based on the attributes of the residual in the pixel domain.
\item \textbf{Cluster-based initialization method. 
} 
This initialization method 
classifies the training data according to attributes of the residual in the frequency domain. As shown in Fig.~\ref{fig:ini_methods}, we first compute the absolute residual between decoded and original patch, which is then transformed by DCT. 
After dimensionality reduction, the DCT coefficients in zigzag-order (called feature vectors) are \js{clustered} to assign a label to each patch. Formally, we can define the Feature Vectors ($FV$) used to cluster as:
\begin{equation}
   FV(X, Y) = Zigzag(DCT(abs(X - Y)))
\end{equation}
where $X \in Q^{W\times H}$ is the decoded frame, $Y \in Q^{W\times H}$ is the original frame, $Q$ is the $GF(2^{8})$ finite field, $abs(\cdot)$ \js{takes} the absolute value of each element, $DCT(\cdot)$ is the DCT \js{operation} while $Zigzag(\cdot)$ flattens the matrix into vector in zigzag scan order. \js{In essence}, $FV(X,Y) \in [a_{i,j}]^{WH\times1}$ where $a_{i,j} \in [0,1]$. Since a lot of DCT coefficients are zero, the dimensionality of the FVs is reduced by t-SNE \cite{maaten2008visualizing}.
\end{itemize}

\noindent We \js{compare} 
these three initialization methods in Section \ref{sec:ex} and visualize \js{their respective classification results of different initialization methods at different iteration steps in Section ~\ref{sec:results_ini_methods} for better understanding their behaviors.}

\subsubsection{Iterative process.} We utilize an iterative training process that updates and refines the classification of training data iteratively to accomplish the training of local CNNs in the {adaptive-switching neural network}. There are two stages in this process: pre-training and iterative update:

\begin{itemize}
\item \textbf{Pre-training.} These three local CNNs are pre-trained separately on three different subsets of the training data \js{with the aim of learning} good initial weights. 
\js{The mean square error (MSE) loss is used during optimization.} \js{To increase the diversity of samples used in these CNNs,} 
the subsets used in pre-training are carefully derived. We first split the training data into five subsets \js{(similar to how it is done in a 5-fold cross-validation)}. 
Then the CNNs of adaptive-switching scheme at QP=37 are pre-trained using three randomly selected subsets \js{from among the five.}

\item \textbf{Iterative update}. Let $F(X_{n};\Theta_{i})$ represent the output of a CNN parameterized by $\Theta_{i}$ for input $X_{n}$. The trained local CNNs for the $(i-1)$-$th$ iteration step are \js{each} parameterized by their respective weights, i.e. $\Theta_{0, i-1}, \Theta_{1, i-1}, \Theta_{2, i-1}$. Meanwhile, the pre-trained global CNN is fixed, with its parameters denoted as $\Theta_{3, i-1}$ at $(i-1)$-$th$ step. \js{In the} performance analysis step shown in Fig.~\ref{fig:iterative_training}, 
a new label for each training patch is generated for each $i$-${th}$ iteration using equation~\ref{eq:training}:

\begin{equation}
\label{eq:training}
  {l}_{n}^{best} = \argmin_{j} PSNR(F(X_{n};\Theta_{j, i-1}),  Y_{n})
\end{equation}

\begin{figure}[t]
\centering
\includegraphics[width=\linewidth]{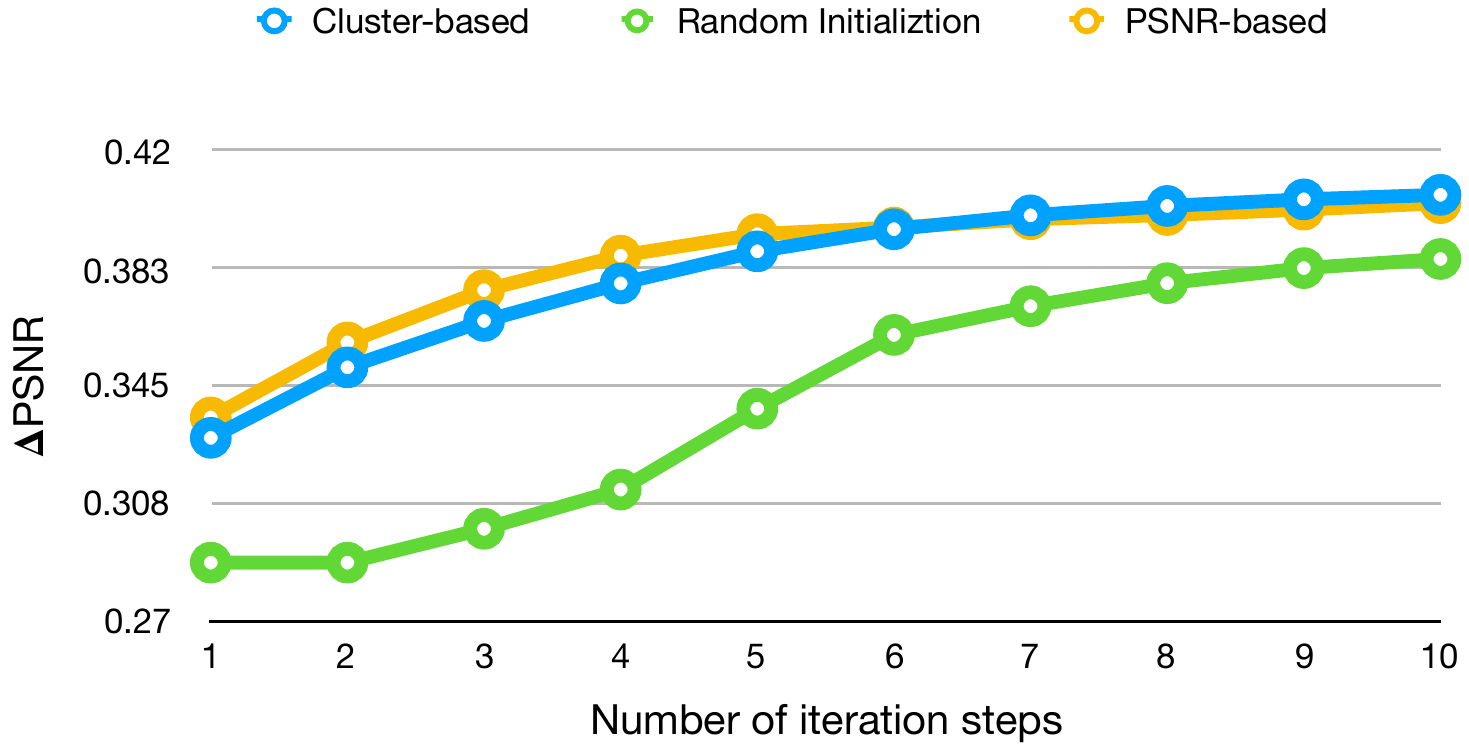}
\caption{Convergence curve for the iterative process in our approach. $x$-axis: number of iteration steps; $y$-axis: gain of \hxy{adaptive-switching neural network} (measured by Y-channel PSNR improvement over HM-16.0 baseline on validation set). Best viewed in color}
\label{fig:convergence}
\end{figure}

\js{for $j \in [0,3]$} where $Y_{n}$ is the corresponding original patch of the input training patch $X_{n}$. After label refinement \js{is completed}, each training patch will be assigned a new label. 
Next, all local CNNs are fine-tuned again on the training set \js{with updated patch labels}. Note that each local CNN is only trained on one particular class of training patches (i.e., subset of patches with a particular patch label value). This procedure is repeated until the gain (PSNR improvement over HM-16.0 baseline) of the \hxy{adaptive-switching neural network} on the validation set \js{converges} or the maximum number of iterations is reached.
\end{itemize}

Although the exact convergence of our iterative process is difficult to analyze due to the inclusion of label refinement, 
our experiments show that the gain of the \hxy{adaptive-switching neural network} \js{stabilizes} within 10 iteration steps for all three initialization methods, which implies the reliability of our approach. From the convergence curves shown in Fig.~\ref{fig:convergence}, we can observe that the average $\Delta$PSNRs of our \hxy{adaptive-switching neural network} 
converged quickly \js{at levels of around} 0.386, 0.403 and 0.406 dB PSNR improvements over HM-16.0 baseline for the random, PSNR-based and cluster-based initialization methods.

We note the following observations about our adaptive-switching scheme:
\begin{enumerate}
\item The local CNNs could focus on a small specific \js{parts of a frame} by way of the patches without
\js{resulting in}
over-fitting since the global CNN can ensure the lower bound of performance.
\item The scheme is independent of the architecture of sub-CNNs utilized 
\item Flags \js{corresponding to the trained CNNs can be easily} binarized and written into bitstreams (2 bits for 4 cases).
\end{enumerate}


\subsection{Test stage}
\label{sec:test}
The framework of our proposed adaptive-switching scheme is shown in Fig.~\ref{fig:framework_b} (encoder side) and Fig.~\ref{fig:overview:d} (decoder side).

\textbf{Encoder side.} 
\js{After} all trained CNNs have been obtained using our proposed iterative training method, we proceed to the encoding procedure. In the encoder side (see Fig.~\ref{fig:framework_b}), an input patch $X_{n}$ is post-processed by all trained CNNs separately. Then the $Flag_{n}$-$th$ CNN is chosen such that the PSNR between $F(X_{n}; \Theta_{Flag_{n}})$ and $Y_{n}$ is smallest across all trained CNNs. 
Formally, we define the flag of the chosen $Flag_{n}$-$th$ CNN as:

\begin{equation}
\label{eq:inf}
  Flag_{n} = \argmin_{j} PSNR(F(X_{n};\Theta_{j}), Y_{n})
\end{equation}
where all notations are the same as in Equation \eqref{eq:training}. These flags indicating the \js{indices} of chosen CNNs for all patches, are written into the bitstream after binarization.

\textbf{Decoder side.} In the decoder side, the flags are \js{first} obtained from the bitstream. For a patch $X_{n}$, the CNN used to accomplish post-processing is chosen according to the corresponding flag $Flag_{n}$. Hence, processing for each decoded frame is achieved by passing \js{each patch} through its selected CNN. Note that we do not know the PSNR between the post-processed patch and original \js{patch} since the original frame is unknown on the decoder side.


\begin{figure}[t]
 \centering
 \includegraphics[width=\linewidth]{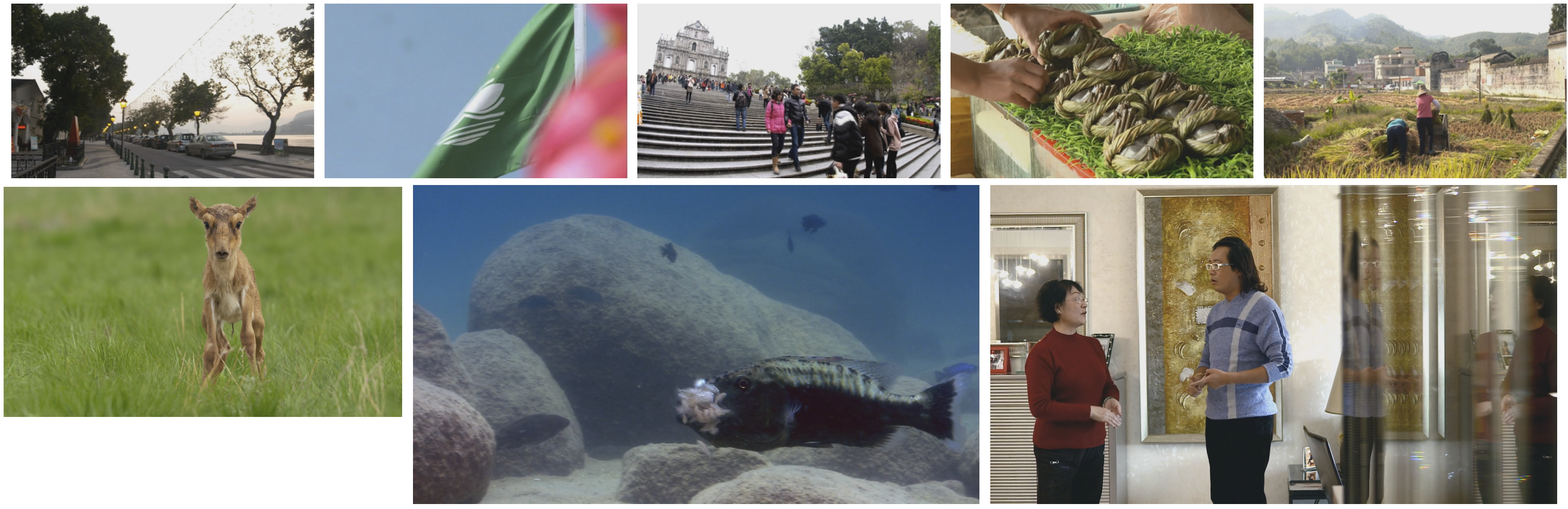}
 \caption{Example snapshots from our training dataset.}
 \label{fig:2b}
\end{figure}

\begin{table*}[t]
\centering

\caption{Comparison of different mask and fusion methods on Y-channel $\Delta$PSNR (dB) over HM-16.0 baseline at QP=27/37 under LP configuration.}
\label{table1}
\begin{tabular}{|c|c|c|c|c|c|c|c|}
\hline
  QP & Class & Sequence & 1-in & 2-in\newline+BM \newline+AF &2-in\newline+MM\newline+CLF&2-in\newline+MM\newline+CEF & 2-in\newline+MM\newline+AF\\
\hline
 \multirow{21}{*}{27/37} & \multirow{4}{*}{A} & Traffic & 0.3497 / 0.3064 & \textbf{0.3908} / 0.3679& 0.3868 / 0.3592 &  0.3600 / 0.3317& 0.3880 / \textbf{0.3943}\\
 \cline{3-8}
 & & PeopleOnStreet& 0.4827 / 0.5592& 0.5180 / 0.6367 & \textbf{0.5225} / 0.6382 & 0.4846 / 0.6087 & 0.4262 / \textbf{0.6410}\\
 \cline{3-8}
 & & Nebuta& 0.4503 / 0.2679& 0.4959 / 0.1994 & \textbf{0.5132} / 0.2491 &0.4707 / 0.3000 & 0.3484 / \textbf{0.3200}\\
 \cline{3-8}
 & & SteamLocomotive & 0.2902 / 0.1857 & 0.3160 / 0.1217 & \textbf{0.3268} / 0.1826 & 0.3113 / 0.1900 & 0.2643 / \textbf{0.2210}\\
 \cline{3-8}
 & & \textbf{Average} & 0.3932 / 0.3298& 0.4302 / 0.3314 & \textbf{0.4373} / 0.3573 &  0.4067 / 0.3576 &0.3567 / \textbf{0.3941} \\
 \cline{2-8}
 & \multirow{5}{*}{B}& Kimono& 0.3859 / 0.3552& \textbf{0.4184} / 0.3932 & 0.4165 / 0.3829 & 0.3860 / 0.3934 & 0.3878 / \textbf{0.4149}\\
 \cline{3-8}
 & & ParkScene& 0.1940 / 0.1693& 0.2232 / 0.1927 & 0.2208 / 0.1883 & 0.2002 / 0.1877 &\textbf{0.2263} / \textbf{0.1994}\\
 \cline{3-8}
  && Cactus& 0.2154 / 0.2288 & 0.2431 / 0.3130 & \textbf{0.2450} / 0.3005 &0.2175 / 0.2704& 0.2412 / \textbf{0.3374}\\
 \cline{3-8}
 & & BQTerrace& 0.1767 / 0.1836&  0.2080 / 0.2950 & 0.2071 / 0.2754 & 0.1889 / 0.2881 &\textbf{0.2429} / \textbf{0.3791}\\
 \cline{3-8}
 & & BasketballDrive& 0.1586 / 0.1918& 0.2010 / 0.3208 & 0.1978 / 0.3076 & 0.1557 / 0.2988 &\textbf{0.2225} / \textbf{0.3518}\\
 \cline{3-8}
 & & \textbf{Average} & 0.2261 / 0.2257& 0.2587 / 0.3029 & 0.2575 / 0.2910 & 0.2297 / 0.2877 &\textbf{0.2641} / \textbf{0.3365} \\
 \cline{2-8}
 & \multirow{4}{*}{C}& RaceHorses& 0.2557 / 0.2594& 0.2793 / \textbf{0.2962} & \textbf{0.2807} / 0.2900 & 0.2530 / 0.2904  & 0.2699 / 0.2943\\
 \cline{3-8}
  & & BQMall& 0.1965 / 0.0954&  0.2698 / 0.2465 & 0.2764 / 0.2280 & 0.2184 / 0.2700 &\textbf{0.3201} / \textbf{0.3554}\\
 \cline{3-8}
  & & PartyScene& 0.1299 / 0.1052& 0.1643 / 0.1728 & 0.1746 / 0.1538 &0.1388 / 0.1890 &\textbf{0.2136} / \textbf{0.2715}\\
 \cline{3-8}
  & & BasketballDrill& 0.2686 / 0.2228& 0.3784 / 0.3401 & 0.3860 / 0.3230 & 0.3093 / 0.3228 &\textbf{0.4490} / \textbf{0.4699}\\
  \cline{3-8}
 & & \textbf{Average} &0.2127 / 0.1707 &0.2730 / 0.2639 & 0.2794 / 0.2487 &0.2299 / 0.2681& \textbf{0.3132} / \textbf{0.3478} \\
 \cline{2-8}
  & \multirow{4}{*}{D}& RaceHorses& 0.3817 / 0.3090& 0.4381 / \textbf{0.4212} & \textbf{0.4424} / 0.4094 & 0.3814 / 0.4130 & 0.3932 / 0.4060\\
 \cline{3-8}
  & & BQSquare& 0.1511 / -0.04& 0.2269 / 0.2218& 0.2216 / 0.1605 & 0.1795 / 0.2442 &\textbf{0.3345} / \textbf{0.4986}\\
 \cline{3-8}
  & & BlowingBubbles& 0.1918 / 0.1250& 0.2463 / 0.2190& 0.2431 / 0.1976 & 0.1993 / 0.2181 &\textbf{0.2593} / \textbf{0.2578}\\
 \cline{3-8}
  & & BasketballPass& 0.2465 / 0.1882& 0.3273 / 0.3510& 0.3233 / 0.3200 & 0.2707 / 0.3555 &\textbf{0.3739} / \textbf{0.4007}\\
  \cline{3-8}
 & & \textbf{Average} &0.2428 / 0.1459&0.3097 / 0.3033 & 0.3076 / 0.2719 & 0.2577 / 0.3077 & \textbf{0.3402} / \textbf{0.3908}\\
 \cline{2-8}
  & \multirow{3}{*}{E}& FourPeople& 0.4619 / 0.4426& 0.5190 / 0.5545 & \textbf{0.5264} / 0.5448 & 0.4551 / 0.5312 & 0.5258 / \textbf{0.6244}\\
 \cline{3-8}
  & & Johnny& 0.2996 / 0.3496& \textbf{0.5190} / \textbf{0.5545} & 0.2604 / 0.4682 & 0.3118 / 0.4531 &0.4285 / 0.5402\\
 \cline{3-8}
  & & KristenAndSara&0.3395 / 0.3927& 0.3967 / 0.5634 & 0.3563 / 0.5230 & 0.3748 / 0.5190 &\textbf{0.4298} / \textbf{0.5942}\\
  \cline{3-8}
 & & \textbf{Average} & 0.3670 / 0.3950& 0.4333 / 0.5317 &  0.3810 / 0.5120 & 0.3806 / 0.5011 & \textbf{0.4614} / \textbf{0.5863}\\
 \cline{2-8}
  & \multicolumn{2}{c|}{\textbf{Average}} & 0.2813 / 0.2449& 0.3322 / 0.3352 & 0.3264 / 0.3251 & 0.2934 / 0.3338 &\textbf{0.3373} / \textbf{0.3986}\\
 \cline{2-8}
\hline
\end{tabular}
\end{table*}


\section{\js{Experiments}}
\label{sec:ex}
\subsection{Dataset}
\js{In order to encourage a more comprehensive validation process,}
we establish a large-scale dataset. The dataset is derived from 600 video clips of various resolutions (i.e. $1920\times1080, 1280\times720, 832\times480$ pixels) with a frame rate of 30 $fps$ for all videos. Fig. \ref{fig:2b} shows some snapshots of the video clips. All raw video clips are encoded by HM-16.0 at Low-delay P at QP=22, 27, 32, and 37. In each raw clip and its compressed clip, we randomly select 3 raw frames and \js{their} corresponding decoded frames to form 3 training frame pairs. For each frame pair, we further divide them into 64$\times$64 \js{non-overlapping} sub-images, resulting in 202,251 sub-image pairs. We use only the luminance channel (Y-channel) for training.

\subsection{\js{Experimental} Settings}
In our experiments, all models are implemented using TensorFlow \cite{tensorflow}. MSE is applied as the loss function for our proposed networks, and \js{it} is formally denoted as:
\begin{equation}
  L(\Theta) = \frac{1}{N} \sum_{n=1}^{N} \Vert F(X_{n};\Theta) - Y_{n} \Vert_{2}^{2}
\end{equation}
\js{with similar notations to earlier equations;} $X_{n}$ is the input compressed frame, $\Theta$ is the learnable parameters of the whole network and $Y_{n}$ is the original frame.

For the partition-aware CNN, we use a mini-batch size of 32. We start with a learning rate of 1e-04, \js{decaying by} a power of 10 at the $20th$ epoch, and terminating at the $40th$ epoch. An individual CNN is trained for each QP. To reduce training time, we first train the CNN at QP=37 from scratch and the other networks at QP=32, 27, 22 are fine-tuned from it.

For all cases, the global CNN of the adaptive-switching scheme is directly trained on the entire training dataset while all local CNNs are trained using the iterative training method. \hxy{Overall, it takes about 26 hours to train a single partition-aware CNN from scratch (on 1 GeForce  GTX  1080Ti  GPU), and 70 hours to train the \hxy{adaptive-switching neural network} from scratch (on 4 GeForce GTX 1080Ti GPUs). }

Note that our adaptive-switching scheme is independent of the architecture of these sub-CNNs, \js{hence allowing the flexibility of} plugging in other existing CNN-based models. In our experiments, we also compare the performance of the proposed adaptive-switching scheme with different CNN architectures.



For the evaluation, we tested our trained model on 20 benchmark sequences (not \js{included} in our training set) from the Common Test Conditions of HEVC \cite{ctc} under the same configuration as with training, Low-delay P \cite{schwarz2005hierarchical}. 
Performance is measured by PSNR improvement ($\Delta$PSNR) and the Rate-distortion performance measured by the Bjontegaard Distortion-rate savings (BD-rate savings, calculated at QP=22, 27, 32, 37) \cite{bd-rate} over the standard HEVC test model HM-16.0 (i.e., HM-16.0 baseline). Basically, a larger PSNR improvement or a larger BD-rate saving indicate that more visual artifacts are reduced.

Note that both in-loop filters (deblocking filter \& SAO filter) are turned on in HM-16.0. \js{For a more in-depth evaluation of our methods, }
performances on other configurations (Low-delay B, Random access, All intra) and color channels are also \js{obtained} for the purpose of comparing \js{against} our methods. Moreover, in order to obtain fair comparison, all methods in Table~\ref{table1} are trained using the same dataset (i.e., our dataset), and evaluated under the same settings.

\begin{table}[t]
\centering
\caption{Comparison of different mask and fusion methods on BD-rate (Y,\%) over HM-16.0 baseline under LP configuration}
\label{tab:RD-mask}
\setlength\tabcolsep{1.5pt}
\begin{tabular}{|c|c|c|M{9mm}|M{9mm}|M{10mm}|M{9mm}|}
\hline
 Class & Sequence & 1-in & 2-in\newline+BM\newline+AF &2-in\newline+MM\newline+CLF&2-in\newline+MM\newline+CEF & 2-in\newline+MM\newline+AF\\
\hline
  \multirow{4}{*}{A} & Traffic &-9.27 &-11.27& -11.25& -10.58 &\textbf{-11.35}\\
 \cline{2-7}
  & PeopleOnStreet&  -9.84& -11.48& \textbf{-11.54} & -10.90 &-10.36 \\
 \cline{2-7}
  & Nebuta& -6.23& -8.03& \textbf{-8.13} & -7.91 & -7.85 \\
 \cline{2-7}
  & SteamLocomotive& -10.22&-11.79 & \textbf{-12.30} & -11.84 & -10.6\\
  \cline{2-7}
  & \textbf{Average} & -8.89 & -10.64 & \textbf{-10.81} & -10.31 & -10.04 \\
 \cline{1-7}
  \multirow{5}{*}{B}& Kimono& -9.49&  \textbf{-11.19} & -11.14 & -10.64 & -10.91 \\
 \cline{2-7}
  & ParkScene& -5.4& -6.74& -6.75 & -6.38& \textbf{-6.92}\\
 \cline{2-7}
  & Cactus& -8.13& -10.50& -10.43 &-9.04 &\textbf{-10.53} \\
 \cline{2-7}
  & BQTerrace& -7.25& -9.45& -9.29 & -6.61&\textbf{-11.07}\\
 \cline{2-7}
  & BasketballDrive& -6.42&-10.06 & -10.14 & -8.85 &\textbf{-11.10}\\
  \cline{2-7}
  & \textbf{Average} & -7.34 & -9.59 & -9.55 & -8.30 & \textbf{-10.10} \\
 \cline{1-7}
  \multirow{4}{*}{C}& RaceHorses& -5.57&-6.6 & \textbf{-6.61} & -6.23& -6.45\\
 \cline{2-7}
  & BQMall& -4.01 & -6.62& -6.75& -6.03&\textbf{-7.62}\\
 \cline{2-7}
  & PartyScene& -2.48& -3.79& -3.85 & -3.55&\textbf{-4.84}\\
 \cline{2-7}
  & BasketballDrill& -5.71& -9.11& -9.20 & -8.12& \textbf{-10.65}\\
  \cline{2-7}
  & \textbf{Average} & -4.44 & -6.53 & -6.60 & -5.98 & \textbf{-7.39} \\
 \cline{1-7}
  \multirow{4}{*}{D}& RaceHorses& -6.66 & -8.29& \textbf{-8.32} & -7.66 & -7.58\\
 \cline{2-7}
  & BQSquare& -2.48& -5.66& -5.58 & -5.09 &\textbf{-8.48}\\
 \cline{2-7}
  & BlowingBubbles& -4.12& -6.08& -6.00 &-5.41 & \textbf{-6.33}\\
 \cline{2-7}
  & BasketballPass& -4.49& -7.02& -6.98 & -6.32& \textbf{-7.73}\\
  \cline{2-7}
  & \textbf{Average} & -4.44 & -6.76 & -6.72 & -6.12 & \textbf{-7.53}\\
 \cline{1-7}
  \multirow{3}{*}{E}& FourPeople& -10.69 & -13.47& -13.55 & -12.69& \textbf{-13.91}\\
 \cline{2-7}
  & Johnny& -10.40& -15.64& -11.67 &-14.17 &\textbf{-17.22} \\
 \cline{2-7}
  & KristenAndSara& -9.5 & -12.94& -11.43 & -12.43&\textbf{-13.78} \\
  \cline{2-7}
  & \textbf{Average} & -10.20 & -14.02 & -12.22 & -13.10 & \textbf{-14.97} \\
 \cline{1-7}
 \multicolumn{2}{|c|}{\textbf{Average}} & -6.92& -9.29& -9.05 & -8.63 & \textbf{-9.76}\\
\hline
\end{tabular}
\end{table}

\begin{table}[t]
\centering
\caption{Comparison of different initialization methods on Y-channel $\Delta$PSNR(dB) over HM-16.0 baseline at QP=37 under LP configuration.}
\label{tab:ini}
\setlength\tabcolsep{3pt}
\begin{tabular}{|c|c|c|c|c|c|c|}
\hline
 \multirow{2}{*}{QP} & \multirow{2}{*}{Class} & \multirow{2}{*}{Sequences} & \multicolumn{3}{c|}{ASN@4$\textit{S}^{\star}$} \\
 \cline{4-6}
 & & & Random & PSNR-based & Cluster-based \\
 \hline
 \multirow{21}{*}{37} & \multirow{4}{*}{A} & Traffic & 0.3048 & 0.3291 & 0.3366\\
 \cline{3-6}
 & & PeopleOnStreet & 0.5640 & 0.5978 & \textbf{0.6139}\\
 \cline{3-6}
 & & Nebuta& 0.2814 & 0.3110 & \textbf{0.3232}\\
 \cline{3-6}
 & & SteamLocomotive &0.1937 & 0.2059 & \textbf{0.2148}\\
  \cline{3-6}
 & & \textbf{Average} &0.3360 & 0.3610 & \textbf{0.3721} \\
 \cline{2-6}
 & \multirow{5}{*}{B}& Kimono & 0.3722 & 0.4020 & \textbf{0.4162}\\
 \cline{3-6}
 & & ParkScene & 0.1752 & 0.1900 & \textbf{0.1929}\\
 \cline{3-6}
  && Cactus  & 0.2473 & 0.2827 & \textbf{0.2994}\\
 \cline{3-6}
 & & BQTerrace & 0.2096 & 0.2965 & \textbf{0.2995}\\
 \cline{3-6}
 & & BasketballDrive & 0.2499 & 0.3003 & \textbf{0.3036}\\
  \cline{3-6}
 & & \textbf{Average}  & 0.2509 & 0.2943 & \textbf{0.3023} \\
 \cline{2-6}
 & \multirow{4}{*}{C}& RaceHorses & 0.2693 & 0.2950 & \textbf{0.2984}\\
 \cline{3-6}
  & & BQMall& 0.2198 & \textbf{0.2851} & 0.2761\\
 \cline{3-6}
  & & PartyScene & 0.1581 & 0.2110 & \textbf{0.2119} \\
 \cline{3-6}
  & & BasketballDrill & 0.2650 & 0.3282 & \textbf{0.3365}\\
   \cline{3-6}
 & & \textbf{Average}  & 0.2281 & 0.2798 & \textbf{0.2807} \\
 \cline{2-6}
  & \multirow{4}{*}{D}& RaceHorses & 0.3725 & 0.4126 & \textbf{0.4170}\\
 \cline{3-6}
  & & BQSquare & 0.2514 & \textbf{0.3529} & 0.3361\\
 \cline{3-6}
  & & BlowingBubbles & 0.1722 & \textbf{0.2093} & 0.2089\\
 \cline{3-6}
  & & BasketballPass & 0.2186 & 0.2926 & \textbf{0.3228}\\
   \cline{3-6}
 & & \textbf{Average}  & 0.2537 & 0.3169 & \textbf{0.3212}\\
 \cline{2-6}
  & \multirow{3}{*}{E}& FourPeople & 0.4372 & 0.4909 & \textbf{0.5393}\\
 \cline{3-6}
  & & Johnny& 0.3748 & 0.4328 & \textbf{0.4645}\\
 \cline{3-6}
  & & KristenAndSara & 0.4556 & 0.5214 & \textbf{0.5314}\\
   \cline{3-6}
 & & \textbf{Average}  & 0.4225 & 0.4817 & \textbf{0.5118}\\
 \cline{2-6}
  & \multicolumn{2}{c|}{\textbf{Average}} &  0.2896 & 0.3374 & \textbf{0.3471}\\
 \cline{2-6}
\hline
\end{tabular}
\end{table}

\subsection{Results of partition-aware CNNs \js{under various strategies}
}
\label{sec:mask_gen}
\subsubsection{Performance of visual artifact reduction}
Table \ref{table1} compares the performances of different mask generation and mask-frame fusion strategies described in Fig.~\ref{fig:mask} and Fig.~\ref{fig:fusin_str} in terms of the Y-channel PSNR improvement over HM-16.0 baseline ($\Delta$PSNR). In Table \ref{table1}, \emph{1-in} represents a single-input baseline 
approach where the mask-flow input is
\js{omitted} 
from the framework (Fig. \ref{fig:cnn_detail}); \emph{2-in+MM+AF} represents our partition-aware CNN using the local mean-based mask and add-based fusion strategy. From Table \ref{table1}, we can observe that:
  \begin{itemize}
      \item When looking at different mask generation strategies, the boundary-based mask strategy (\js{2-in+BM+AF}) provides 0.33 dB PSNR improvement that is similar to the local mean-based mask (\js{2-in+MM+AF}) at QP=27. However, its performance at QP=37 is lower than \emph{2-in+MM+AF} by 0.06 dB. This is because only marking boundary pixels in a mask is slightly less effective in highlighting the partition modes in a frame across different QPs. 

      \item As for mask-frame fusion strategies, the add-fusion strategy (\js{2-in+MM+AF}) can obtain large PSNR improvements of 0.40 dB at QP=37 and 0.34 dB at QP=27. This shows the effectiveness of the proposed fusion strategy. Comparatively, the concatenate-based late fusion (2-in+MM+CLF)  and early-fusion (\js{2-in+MM+CEF}) strategies obtain smaller gains at both QP=27, 37. This is probably because these fusion strategies are less compatible with the CNN model used in this paper. 

      \item The best performance is obtained when using local mean-based mask and add-fusion (\js{2-in+MM+AF}), which can obtain over 0.15 dB at QP=37 improvement over single-input method. Similar results can be found at QP=27. This indicates that when strategies are properly selected, introducing partition information is indeed useful to reduce the visual artifacts of compressed videos.

      \item Our baseline single-input method (\emph{1-in}) can also obtain satisfactory results. This implies that the baseline CNN model used in our approach is effective in handling the visual information of the input decoded frames.
  \end{itemize}
 \subsubsection{Rate-distortion performance} We also compare the BD-rate saving  of different mask and fusion methods over HM-16.0 in Table \ref{tab:RD-mask}. \js{Comparisons between these methods can be summarized as follows:} 
 (1) The local mean-based mask (\emph{2-in+MM+AF}) achieves 0.47\% BD-rate saving more than the boundary-based mask strategy (\emph{2-in+BM+AF}); (2) The concatenate-based late-fusion (\emph{2-in+MM+CLF}) and early-fusion (\emph{2-in+MM+CEF}) strategies obtain 9.05\% and 8.63\% BD-rate saving, which both are lower than the local mean-based mask (\emph{2-in+MM+AF}); (3) Our approach using local mean-based mask and add-fusion (\emph{2-in+MM+AF}) is able to achieve up to 9.76\% BD-rate saving over all test sequences. This again validates the effectiveness of introducing partition information when strategies are properly selected. These observations are also consistent with the PSNR improvement measure.

\begin{figure}[t]
 \centering

\includegraphics[width=\linewidth]{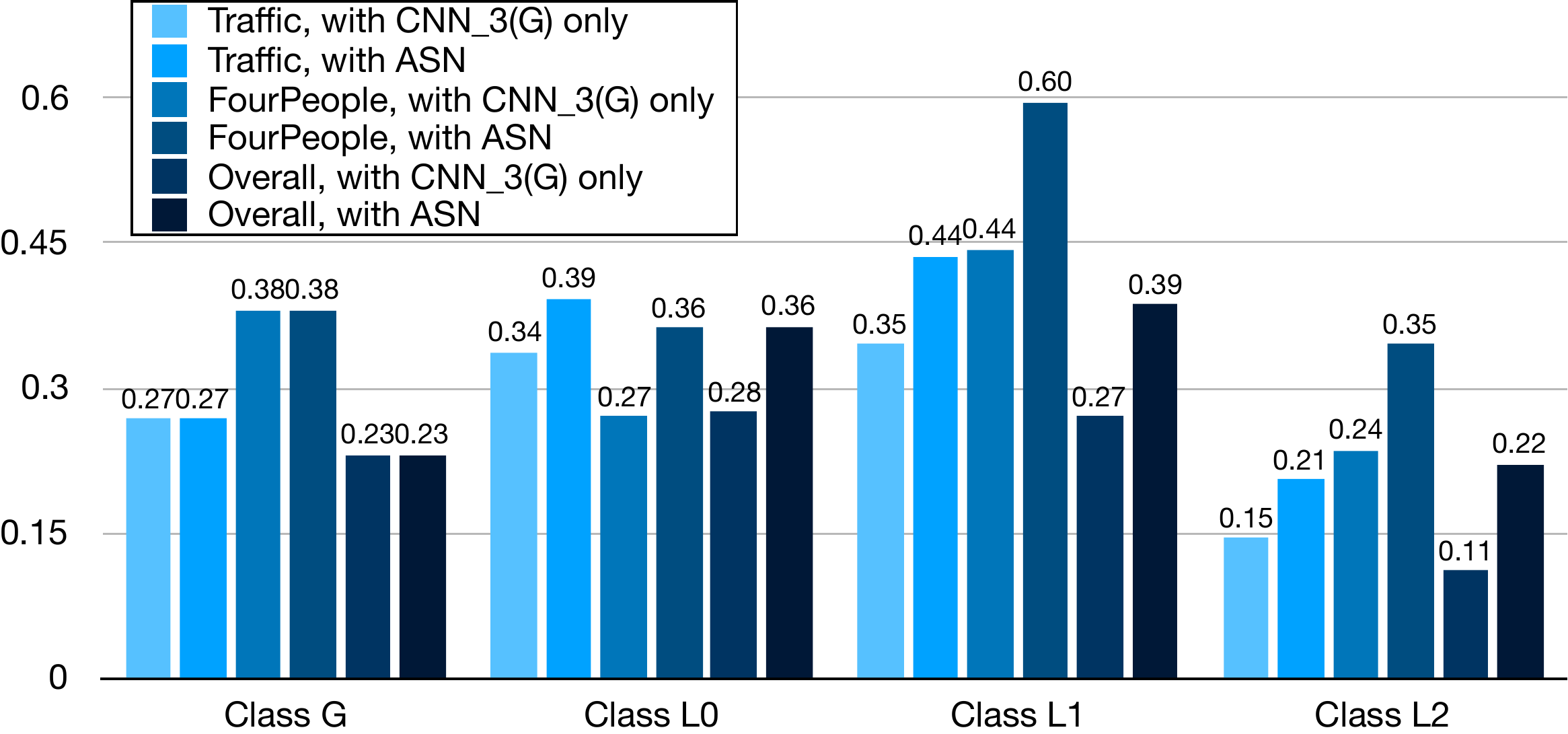}
 \caption{Average Y-channel PSNR improvement over HM-16.0 baseline of different classes of patches post-processed by ASN@4$\textit{S}^{\star}$ and its CNN\_3(G) only at QP=37. Best viewed in color.}
 \label{fig:average_PSNR}
\end{figure}


\begin{figure}[ht]
 \centering
\includegraphics[width=\linewidth]{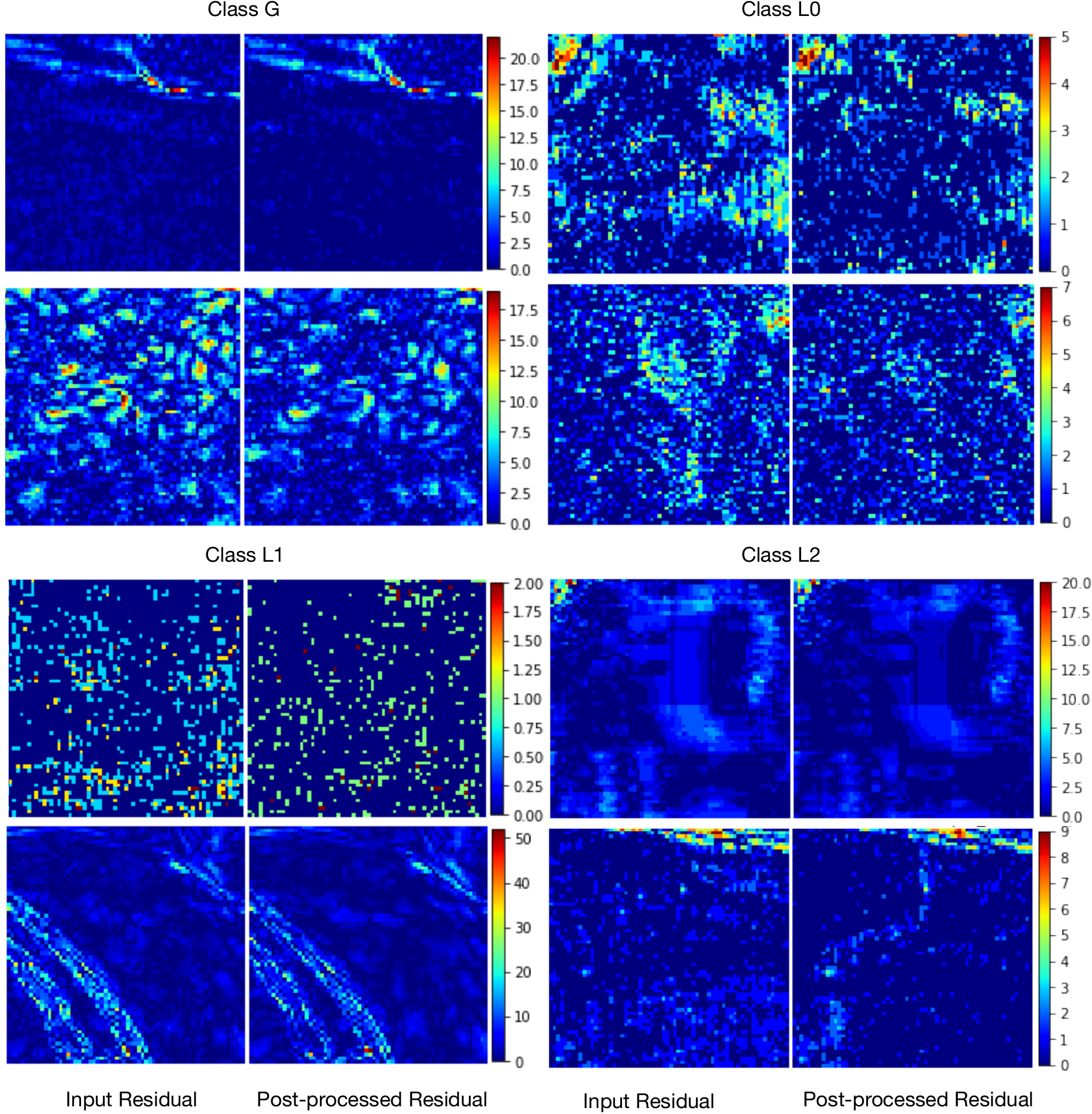}
 \caption{Examples of different classes of patches and corresponding for ASN@4$\textit{S}^{\star}$ with Cluster-based initialization method at the final iteration step. Best viewed in color.}
 \label{fig:examples_of_patches}
\end{figure}

\begin{table}
\centering
\caption{notations of the CNNs used in the experiments of our adaptive-switching scheme}
\label{tab:notation}
\begin{tabular}{|c|M{4cm}|c|}
\hline
     Original name & Description & Notation  \\
     \hline
     VRCNN~\cite{vrcnn} & A shallow model proposed by~\cite{vrcnn}  & $\textit{S}$ \\
     \hline
     VRCNN+MM+AF & A partition-aware shallow model, which integrates our partition-aware-based approach into the existing VRCNN method & $\textit{S}^{\star}$ \\
     \hline
     1-in & Our proposed single-input baseline without mask-flow input & $\textit{D}$ \\
     \hline
     2-in+MM+AF & A partition-aware deep model of 1-in& $\textit{D}^{\star}$ \\
\hline
\end{tabular}
\end{table}
\subsection{Results of our adaptive-switching scheme }
As shown in Table~\ref{tab:notation}, the following architectures are included in our \js{comparative} experiments: (1) A shallow model VRCNN~\cite{vrcnn} denoted by \textit{S}; (2) A deep model, \js{which is} our proposed \emph{1-in} denoted by \textit{D}; (3) A partition-aware shallow model, VRCNN+MM+AF, which integrates our proposed partition \js{awareness} into the existing VRCNN~\cite{vrcnn} method, which is denoted by $\textit{S}^{\star}$, and (4) Our partition-aware deep model, \js{2-in+MM+AF} which is denoted by $\textit{D}^{\star}$. 
\subsubsection{\js{Comparison of various} initialization methods}\label{sec:results_ini_methods}

 Table \ref{tab:ini} shows the Y-channel PSNR improvement \js{over HM-16.0 baseline for the three different initialization methods described in Section \ref{sec:iterative_training}.} \js{To demonstrate this, we choose} the ASN@4$\textit{S}^{\star}$ approach, which is our \hxy{adaptive-switching neural network} (ASN) \js{with four sub-CNNs that are of} the partition-aware shallow model $\textit{S}^{\star}$. We make the following observations:

\begin{itemize}
  \item Our approach with Cluster-based and PSNR-based initialization methods outperforms the Random method over all test sequences at QP=37. 
  \js{Obviously,} the random initialization method is not able to provide a reasonable initial classification \js{and thus resulted in a} 
  lower performance compared with the other two methods.
  \item Our approach with Cluster-based initialization slightly outperforms that with the PSNR-based initialization by only 0.01dB. \js{Specifically,} 
  the PSNR \js{measure} is proportional to the MSE loss function used and \js{this} 
  led to a \js{reasonably competitive} result from the PSNR-based method. However, the Cluster-based method \js{may be able to better} 
  capture the patch features in the frequency domain, achieving \js{marginally} better initial classification. Therefore, we \js{decided to choose this} initialization method for the rest of experiments in this paper.
\end{itemize}

\js{To provide a control experiment on our choice of approach, we demonstrate that the iterative training process plays a major role for the case of our adaptive switching network, and this is consistent across all individual patch classes. } Fig.~\ref{fig:average_PSNR} shows the average Y-channel PSNR improvement of each class of patches post-processed by ASN@4$\textit{S}^{\star}$ with Cluster-based initialization method, \js{and how it matches up against a single global} \textit{CNN\_3(G)}. \js{Each group of vertical bars indicate the class of patches to be post-processed, e.g.} \textit{Class G} means that this class of patches are post-processed by \textit{CNN\_3(G)} 
on the decoder side when ASN is applied. As we can observe, the PSNR improvement achieved by ASN@4$\textit{S}^{\star}$ is larger than that achieved by its \textit{CNN\_3(G)} for each patch class for the \textit{Traffic} sequence, \textit{FourPeople} sequence and overall across all benchmark sequences. 
Fig.~\ref{fig:examples_of_patches} shows some examples of each class of patches. For each patch, we display the heatmap of residual between decoded patch and its ground truth (called input residual), and the heatmap of residual between post-processed patch and its ground truth (called post-processed residual). We can see that the post-processed residual is \js{less intensive}
than its corresponding input residual in all patch classes.


\begin{table*}[t]
\centering
\caption{Comparison of ASN with different CNN architectures on Y-channel $\Delta$PSNR at QP=27, 37 under LP configuration.}
\label{tab:difCNN}
\begin{tabular}{|c|c|c|c|c|c|c||c|c|c|c|c|}
\hline
 QP & Class & Sequence & ASN@4\textit{S} & ASN@4$\textit{S}^{\star}$ & ASN@4\textit{D} & ASN@4$\textit{D}^{\star}$ & ASN@$\textit{D}^{\star}$+3$\textit{S}^{\star}$\\
\hline
   \multirow{21}{*}{27  /  37} & \multirow{4}{*}{A} & Traffic  & 0.3192 / 0.2907 & 0.3581 / 0.3366 & 0.4269 / 0.4437 & \textbf{0.4558} / \textbf{0.4579} & 0.4448 / 0.4106 \\
 \cline{3-8}
 & & PeopleOnStreet & 0.4234 / 0.4670 & 0.5004 / 0.6139 & 0.5611 / 0.6995 & \textbf{0.6088} / \textbf{0.7286} & 0.6010 / 0.6781\\
 \cline{3-8}
 & & Nebuta  & 0.4148 / 0.2401 & 0.5123 / 0.3232 & 0.5259 / 0.3594 & 0.5727 / \textbf{0.3829} & \textbf{0.5736} / 0.3321\\
 \cline{3-8}
 & & SteamLocomotive & 0.2831 / 0.1844 & 0.3324 / 0.2148 & 0.3484 / 0.2639 & \textbf{0.3742} / \textbf{0.2832} & 0.3728 / 0.2515\\
  \cline{3-8}
 & & \textbf{Average} & 0.3601 / 0.2955 & 0.4258 / 0.3721 & 0.4656 / 0.4416 & \textbf{0.5029} / \textbf{0.4631} & 0.4980 / 0.4181\\
 \cline{2-8}
 & \multirow{5}{*}{B}& Kimono& 0.3377 / 0.3182 & 0.3938 / 0.4162 & 0.4496 / 0.4835 & \textbf{0.4860} / \textbf{0.4998} & 0.4802 / 0.4742\\
 \cline{3-8}
 & & ParkScene& 0.1770 / 0.1684 & 0.2064 / 0.1929 & 0.2342 / 0.2174 & \textbf{0.2690} / \textbf{0.2344} & 0.2649 / 0.2144\\
 \cline{3-8}
  && Cactus& 0.1942 / 0.2315 & 0.2307 / 0.2994 & 0.2609 / 0.3882 & \textbf{0.2862} / \textbf{0.4044} & 0.2776 / 0.3596\\
 \cline{3-8}
 & & BQTerrace& 0.1859 / 0.2232 & 0.2203 / 0.2995 & 0.2553 / 0.4154 & \textbf{0.2878} / \textbf{0.4328} & 0.2768 / 0.3310\\
 \cline{3-8}
 & & BasketballDrive& 0.1370 / 0.2100 & 0.1813 / 0.3036 & 0.2425 / 0.4112 & \textbf{0.2723} / \textbf{0.4297} & 0.2477 / 0.3708\\
  \cline{3-8}
 & & \textbf{Average}& 0.2064 / 0.2303 & 0.2465 / 0.3023 & 0.2885 / 0.3831 & \textbf{0.3203} / \textbf{0.4002} & 0.3094 / 0.3500 \\
 \cline{2-8}
 & \multirow{4}{*}{C}& RaceHorses & 0.2322 / 0.2361 & 0.2734 / 0.2984 & 0.2932 / 0.3347 & \textbf{0.3267} / \textbf{0.3583} & 0.3126 / 0.3230\\
 \cline{3-8}
  & & BQMall& 0.1903 / 0.2067 & 0.2443 / 0.2761 & 0.3376 / 0.3837 & \textbf{0.3647} / \textbf{0.4200} & 0.3240 / 0.3293\\
 \cline{3-8}
  & & PartyScene& 0.1295 / 0.1404 & 0.1774 / 0.2119 & 0.2166 / 0.2881 & \textbf{0.2463} / \textbf{0.3120} & 0.2370 / 0.2385\\
 \cline{3-8}
  & & BasketballDrill& 0.2053 / 0.2434 & 0.3145 / 0.3365 & 0.4905 / 0.5348 & \textbf{0.5211} / \textbf{0.5549} & 0.4896 / 0.4455\\
   \cline{3-8}
 & & \textbf{Average} & 0.1894 / 0.2067 & 0.2524 / 0.2807 & 0.3345 / 0.3853 & \textbf{0.3647} / \textbf{0.4113} & 0.3408 / 0.3341 \\
 \cline{2-8}
  & \multirow{4}{*}{D}& RaceHorses & 0.3229 / 0.3278 & 0.4139 / 0.4170 & 0.4561 / 0.456 & \textbf{0.4974} / \textbf{0.4886} & 0.4844 / 0.4532\\
 \cline{3-8}
  & & BQSquare& 0.1543 / 0.2232 & 0.2460 / 0.3361 & 0.3387 / \textbf{0.5286} & \textbf{0.3563} / 0.5196 & 0.3223 / 0.3621\\
 \cline{3-8}
  & & BlowingBubbles& 0.1515 / 0.1646 & 0.2040 / 0.2089 & 0.2787 / 0.2754 & \textbf{0.3126} / \textbf{0.2960} & 0.2987 / 0.2456\\
 \cline{3-8}
  & & BasketballPass& 0.1932 / 0.2230 & 0.2681 / 0.3228 & 0.4005 / 0.4558 & \textbf{0.4324} / \textbf{0.4865} & 0.4063 / 0.3934\\
   \cline{3-8}
 & & \textbf{Average} & 0.2055 / 0.2346 & 0.2830 / 0.3212 & 0.3685 / 0.4289 & \textbf{0.3997} / \textbf{0.4477} & 0.3779 / 0.3636\\
 \cline{2-8}
  & \multirow{3}{*}{E}& FourPeople& 0.4001 / 0.4207 & 0.4763 / 0.5393 & 0.5942 / 0.7027 & \textbf{0.6247} / \textbf{0.7350} & 0.5981 / 0.6496\\
 \cline{3-8}
  & & Johnny & 0.2862 / 0.3624 & 0.3428 / 0.4645 & 0.4542 / 0.6493 & \textbf{0.4913} / \textbf{0.6515} & 0.4653 / 0.5509\\
 \cline{3-8}
  & & KristenAndSara& 0.3517 / 0.4488 & 0.3967 / 0.5314 & 0.4686 / 0.7162 & \textbf{0.4912} / \textbf{0.7290} & 0.4733 / 0.6298\\
   \cline{3-8}
 & & \textbf{Average} & 0.3460 / 0.4106 & 0.4052 / 0.5118 & 0.5056 / 0.6894 & \textbf{0.5357} / \textbf{0.7052} & 0.5122 / 0.6101\\
 \cline{2-8}
  & \multicolumn{2}{c|}{\textbf{Average}}& 0.2545 / 0.2665 & 0.3146 / 0.3471 & 0.3817 / 0.4504 & \textbf{0.4139} / \textbf{0.4702} & 0.3975 / 0.4022\\
 \cline{2-8}
\hline
\end{tabular}
\end{table*}

\begin{table}[t]
\centering
\caption{Comparison of ASN with different CNN architectures on BD-rate (Y,\%) over HM-16.0 baseline under LP configuration.}
\setlength\tabcolsep{1.5pt}
\label{tab:RD-difCNN}
\begin{tabular}{|c|c|M{10mm}|M{10mm}|M{10mm}|M{10mm}|M{10mm}|}
\hline
 Class & Sequence & ASN@\newline 4\textit{S} & ASN@\newline 4$\textit{S}^{\star}$ & ASN@\newline 4\textit{D} & ASN@\newline 4$\textit{D}^{\star}$ & ASN@\newline $\textit{D}^{\star}$+3$S^{\star}$\\
\hline
  \multirow{4}{*}{A} & Traffic  & -8.83 & -9.86 & -11.73 & \textbf{-12.54} & -12.14\\
 \cline{2-7}
 & PeopleOnStreet  & -9.33 & -10.79 & -11.54 & -12.51 & \textbf{-12.66}\\
 \cline{2-7}
 & Nebuta  & -7.14 & -8.84 & -8.58 & \textbf{-12.63} & -9.61\\
 \cline{2-7}
 & SteamLocomotive & -10.43 & -12.16 & -12.90 & \textbf{-14.39} & -13.71\\
 \cline{2-7}
 & \textbf{Average} & -8.93 & -10.41 & -11.19 & \textbf{-13.02} & -12.03 \\
 \hline
  \multirow{5}{*}{B}& Kimono  & -9.26 & -10.58 & -11.74 & \textbf{-12.60} & -12.48\\
 \cline{2-7}
  & ParkScene & -5.25 & -6.04 & -6.64 & \textbf{-7.89} & -7.47\\
 \cline{2-7}
  & Cactus  & -8.04 & -9.64 & -10.75 & -11.27 & \textbf{-11.64}\\
 \cline{2-7}
  & BQTerrace & -7.97 & -9.66 & -11.29 & \textbf{-12.35} & -11.85\\
 \cline{2-7}
  & BasketballDrive & -6.76 & -8.89 & -11.04 & -11.58 & \textbf{-11.68}\\
   \cline{2-7}
 & \textbf{Average}  & -7.46 & -8.96 & -10.29 & \textbf{-11.14} & -11.02 \\
 \hline
  \multirow{4}{*}{C}& RaceHorses & -5.45 & -6.39 & -6.61 & \textbf{-7.51} & -7.20\\
 \cline{2-7}
   & BQMall & -4.48 & -5.84 & -7.43 & \textbf{-8.15} & -7.55\\
 \cline{2-7}
   & PartyScene  & -2.86 & -3.90 & -4.63 & \textbf{-5.14} & -5.04\\
 \cline{2-7}
   & BasketballDrill  & -5.30 & -7.61 & -11.08 & \textbf{-12.24} & -11.19\\
    \cline{2-7}
 & \textbf{Average}  & -4.52 & -5.93 & -7.44 & \textbf{-8.26} & -7.75 \\
 \hline
   \multirow{4}{*}{D}& RaceHorses  & -6.16 & -7.70 & -8.08 & \textbf{-9.24} & -8.82\\
 \cline{2-7}
   & BQSquare  & -3.99 & -6.07 & -8.24 & \textbf{-8.47} & -7.66\\
 \cline{2-7}
   & BlowingBubbles & -3.65 & -4.84 & -6.42 & \textbf{-7.33} & -6.83\\
 \cline{2-7}
   & BasketballPass  & -3.70 & -5.14 & -7.52 &\textbf{ -8.18} & -7.70\\
    \cline{2-7}
 & \textbf{Average}  & -4.37 & -5.94 & -7.56 & \textbf{-8.31} & -7.75 \\
 \hline
   \multirow{3}{*}{E}& FourPeople  & -9.58 & -11.41 & -14.00 & \textbf{-14.44} & -14.34\\
 \cline{2-7}
 & Johnny & -10.61 & -12.73 & -16.72 &\textbf{ -17.35} & -17.01\\
 \cline{2-7}
   & KristenAndSara  & -9.90 & -11.45 & -13.62 & -13.64 & \textbf{-13.75}\\
    \cline{2-7}
 & \textbf{Average}& -10.03 & -11.86 & -14.78 & \textbf{-15.14} & -15.03 \\
 \hline
   \multicolumn{2}{|c|}{\textbf{Average}}& -6.94 & -8.48 & -10.03 & \textbf{-10.97} & -10.52\\
 \hline
 \hline
   \multicolumn{2}{|M{30mm}|}{\textbf{Average(corresp. single-CNN-based method)}}& -3.57 & -6.88 & -6.92 & \textbf{-9.76} & $\backslash$ \\
\hline
\end{tabular}
\end{table}

\begin{figure}[t]
\centering
\subfloat[ASN@4\textit{S}]{\includegraphics[width=0.45\linewidth]{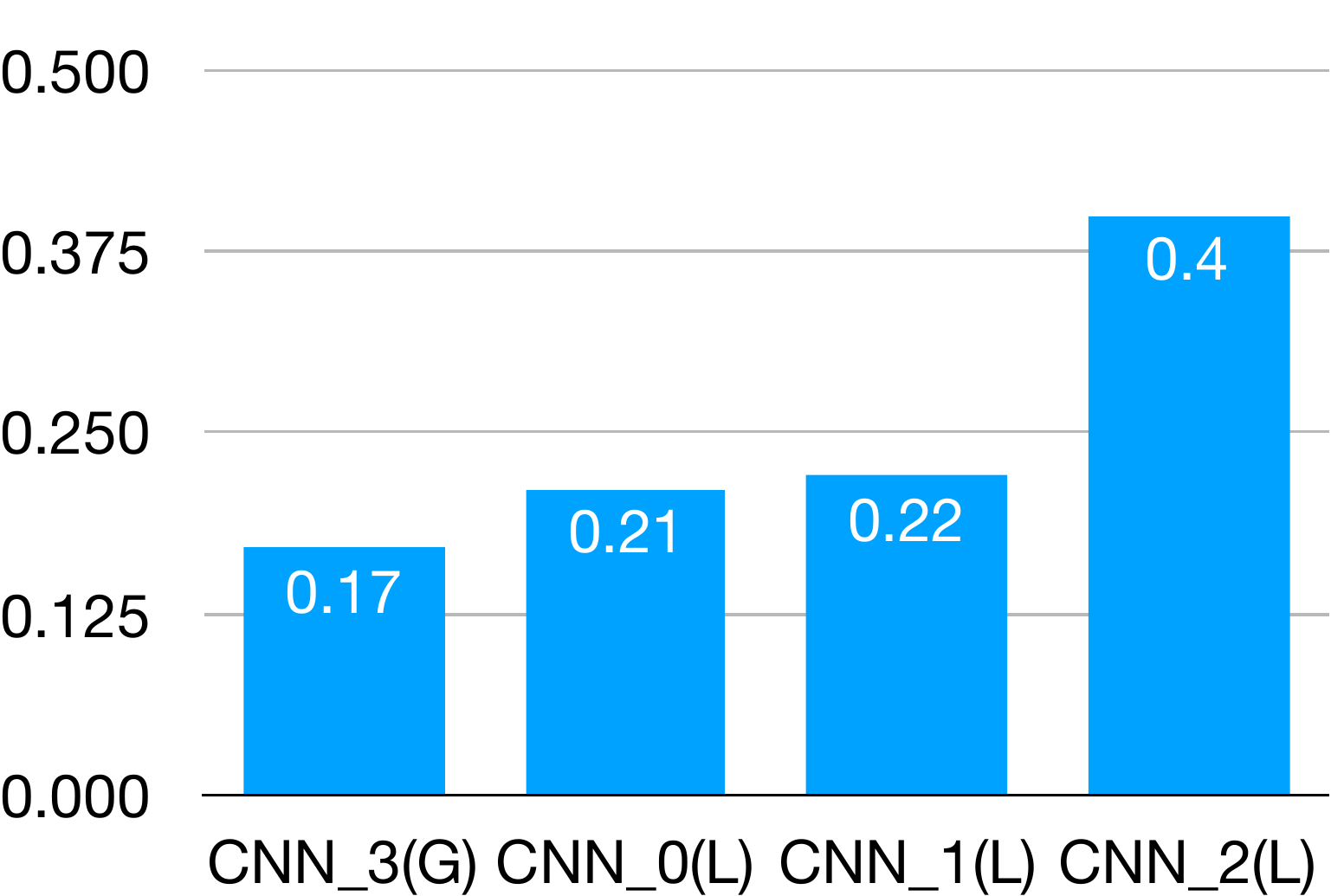}\label{fig:usage_a}}
\hfill
\subfloat[ASN@4\textit{D}]{\includegraphics[width=0.45\linewidth]{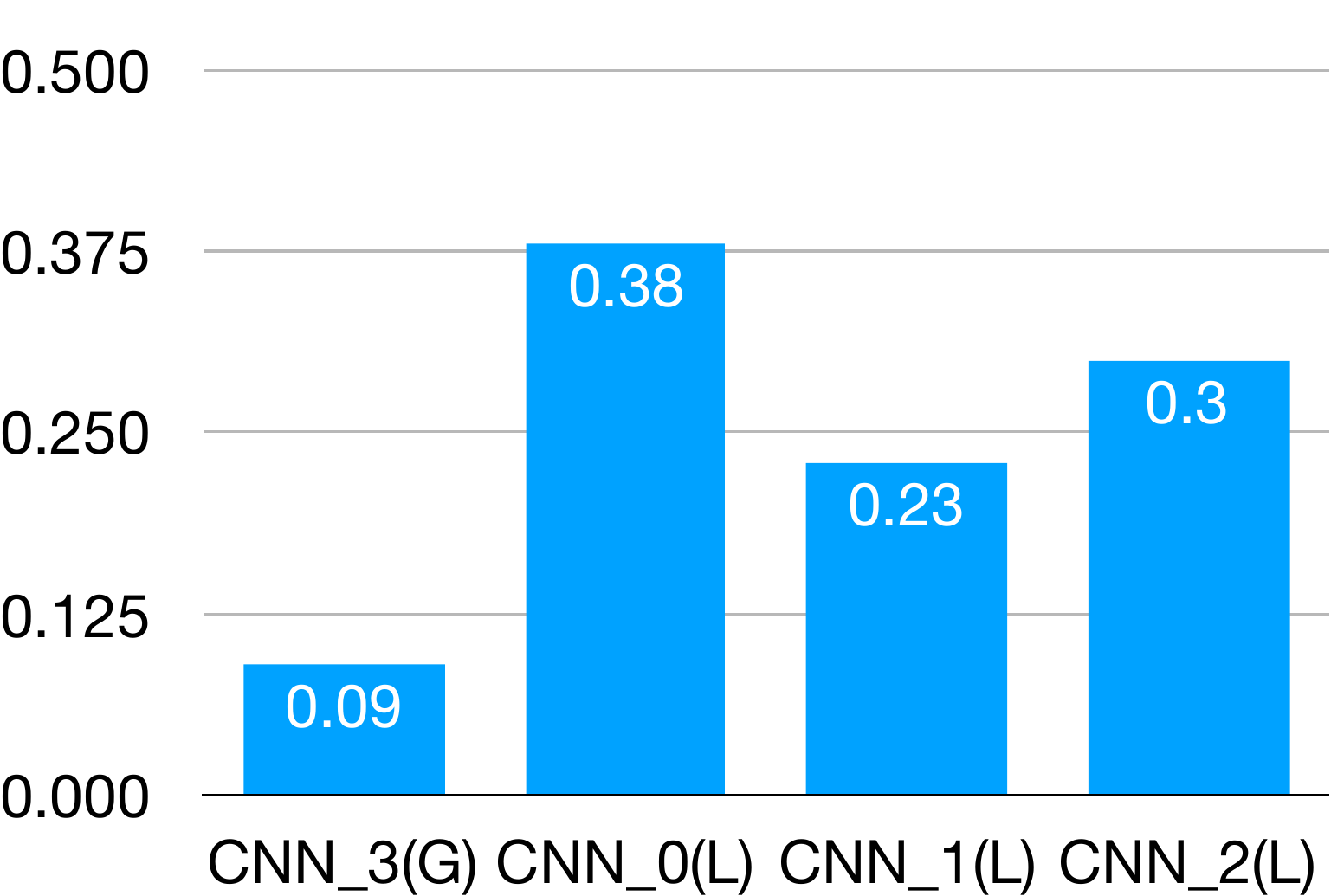}\label{fig:usage_b}}
\hfill
\subfloat[ASN@4$\textit{S}^{\star}$]{\includegraphics[width=0.45\linewidth]{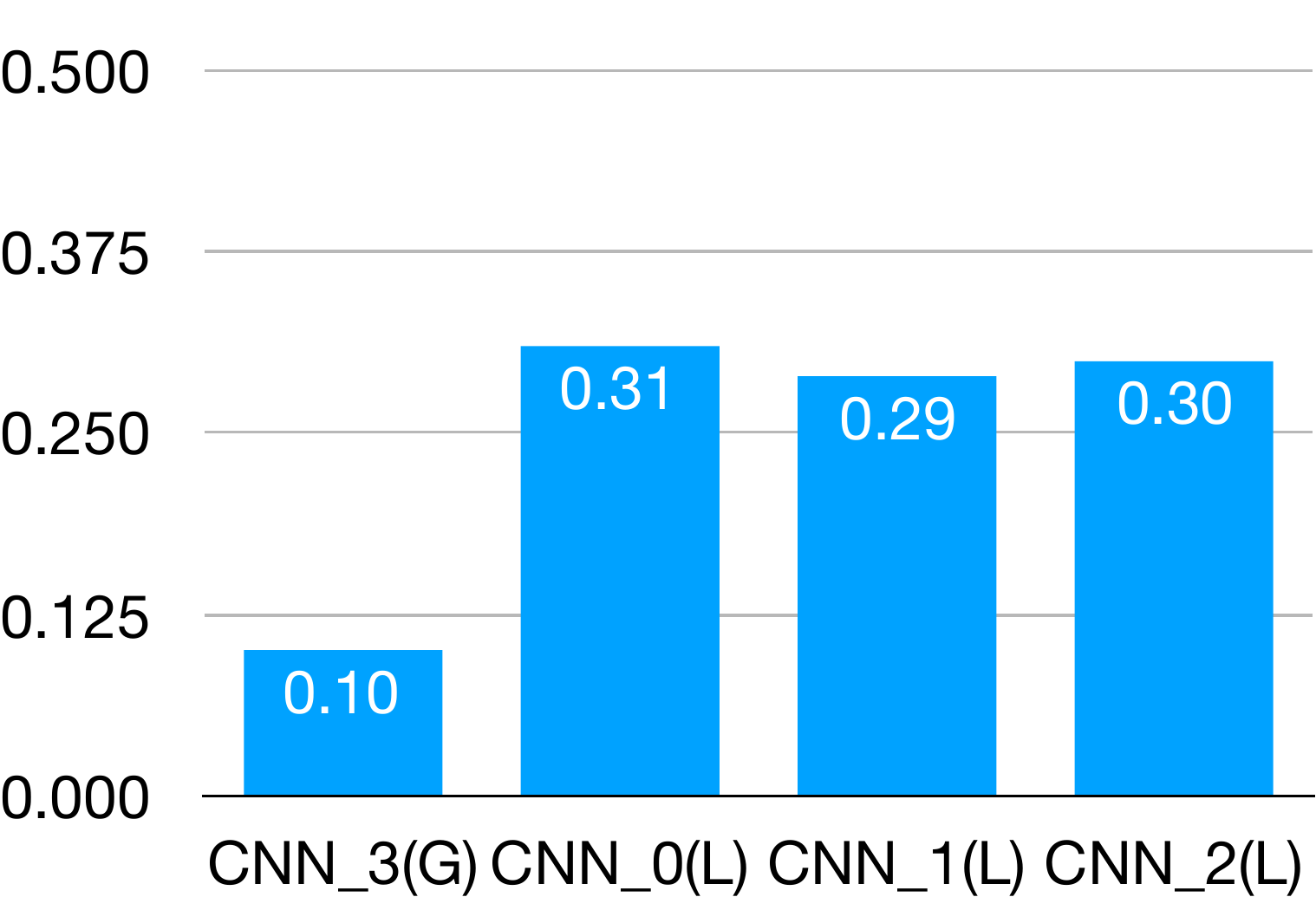}\label{fig:usage_c}}
\hfill
\subfloat[ASN@4$\textit{D}^{\star}$]{\includegraphics[width=0.45\linewidth]{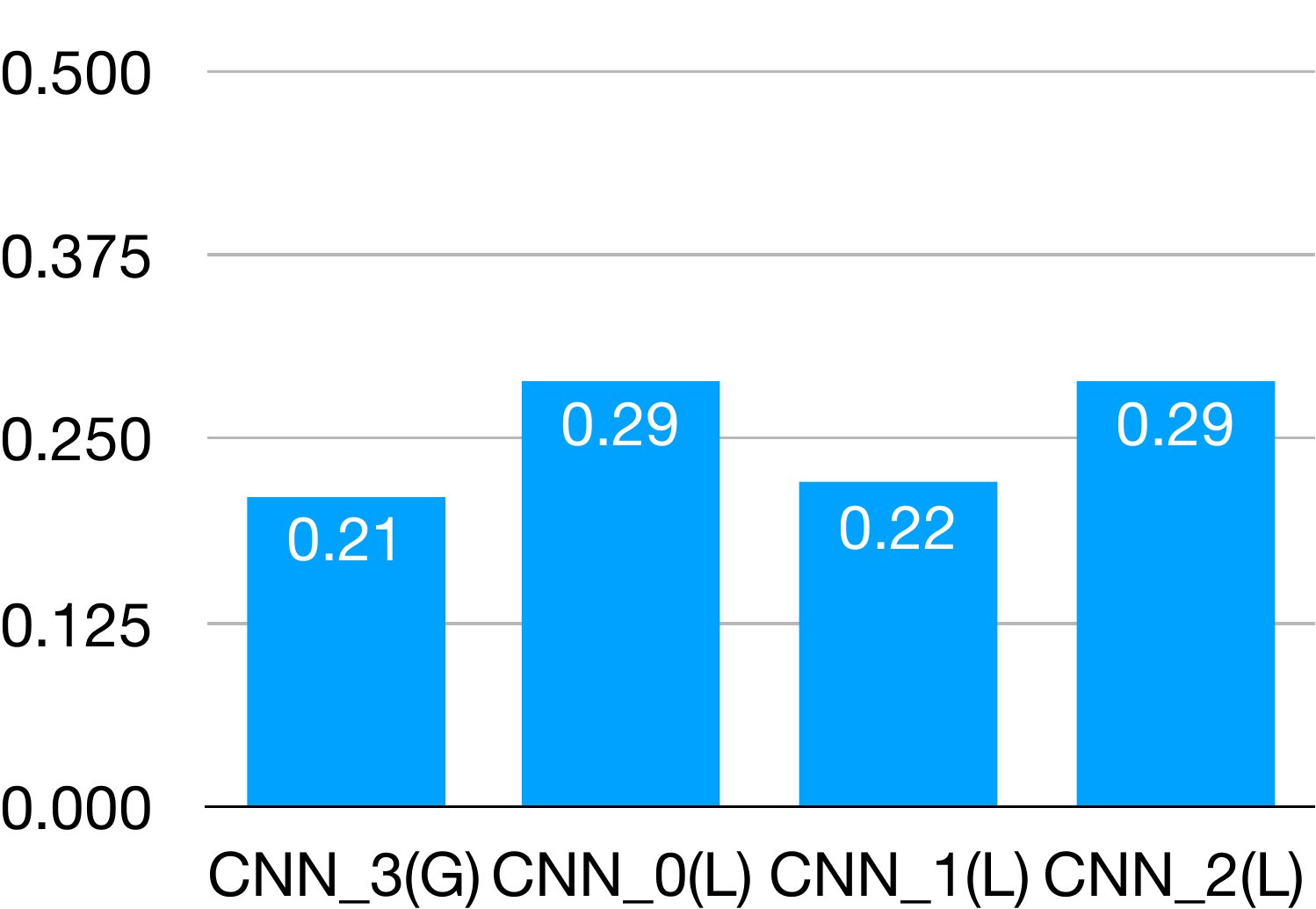}\label{fig:usage_d}}
\hfill
\caption{Usage rate for adaptive-switching scheme with different CNN architectures. $x$-axis: model name in ASN; $y$-axis: usage rate during test at QP=37.}
\label{fig:usage_rate}
\end{figure}

\begin{table*}[ht]
\centering
\caption{\hxy{Comparison of different methods on BD-rate (\%) saving over HM-16.0 baseline under different configurations.}}
\label{table2}
\setlength\tabcolsep{1.0pt}
\begin{tabular}{|c|c|c|c|c|c|c|c|c|c|c|c|c|c|c|c|c|c|c|c|c|c|c|}
\hline
 \multirow{2}{*}{Conf.} & \multirow{2}{*}{Seq.} & \multicolumn{3}{c|}{VRCNN~\cite{vrcnn}} & \multicolumn{3}{c|}{QECNN-P\cite{qecnn}} & \multicolumn{3}{c|}{DRN\cite{DRN2019}} & \multicolumn{3}{c|}{VRCNN+MM+AF} & \multicolumn{3}{c|}{DRN+MM+AF} & \multicolumn{3}{M{21.6mm}|}{Our 2-in+MM+AF \newline ($\textit{D}^{\star}$)} & \multicolumn{3}{c|}{Our ASN@4$\textit{D}^{\star}$}\\
 \cline{3-23}
 & & Y & U & V & Y & U & V & Y & U & V & Y & U & V & Y & U & V & Y & U & V & Y & U & V \\
\hline
  \multirow{6}{*}{LP} & Class A & -7.10 & -2.41 & -1.97 & -8.43 & -3.21 & -2.78 & -8.66 & -3.58 & -3.11 & -9.00 & -3.89 & -3.48 & -8.80 & -4.42 & -4.10 & -10.04 & -6.04 & -5.72 & -12.02 & -6.63 & -6.33\\
 \cline{2-23}
  & Class B& -4.57 & -4.13 & -5.44 & -6.39 & -5.25 & -7.21 & -6.65 & -5.91 & -7.19 & -7.28 & -6.52 & -8.30 & -8.12 & -6.99 & -8.96 & -10.10 & -9.48 & -11.94 & -11.14 & -10.64 & -13.56\\
 \cline{2-23}
  & Class C& -0.21 & -2.73 & -4.12 & -2.81 & -3.92 & -5.59 & -3.55 & -4.45 & -6.19 & -4.05 & -5.11 & -7.03 & -4.85 & -5.34 & -7.06 & -7.39 & -8.56 & -11.07 & -8.26 & -9.66 & -12.51\\
 \cline{2-23}
  & Class D& 0.49 & -1.88 & -2.66 & -2.52 & -2.90 & -3.70 & -3.64 & -3.32 & -4.37 & -4.16 & -3.74 & -4.86 & -5.47 & -4.38 & -5.73 & -7.53 & -6.76 & -8.36 & -8.31 & -8.23 & -8.21\\
  \cline{2-23}
  & Class E& -7.11 & -11.25 & -12.24 & -9.47 & -13.21 & -14.02 & -9.37 & -13.16 & -14.00 & -10.79 & -14.42 & -15.08 & -11.97 & -15.08 & -15.51 & -14.97 & -17.70 & -17.75 & -15.14 & -18.52 & -18.41\\
  \cline{2-23}
  & \textbf{Average}& -3.57 & -4.13 & -4.95 & -5.77 & -5.30 & -6.32 & -6.24 & -5.72 & -6.63 & -6.88 & -6.34 & -7.41 & -7.65 & -6.84 & -7.94 &-9.76 & -9.30 & -10.68 & \textbf{-10.97} & \textbf{-10.34} & \textbf{-11.56}\\
    \hline
    \hline
  \multirow{6}{*}{LB} & Class A& -4.83 & -1.13 & -0.76 & -5.85 & -1.80 & -1.45 & -6.13 & -2.27 & -1.89 & -6.43 & -2.56 & -2.23 & -6.96 & -3.03 & -2.88 & -7.74 & -5.17 & -4.88 & -9.28 & -5.57 & -5.31\\
 \cline{2-23}
  & Class B& -2.26 & -2.68 & -3.90 & -3.92 & -3.68 & -5.57 & -4.30 & -4.45 & -5.59 & -4.73 & -5.00 & -6.64 & -5.49 & -5.40 & -7.22 & -7.58 & -8.33 & -10.68 & -8.33 & -9.21 & -11.95\\
 \cline{2-23}
  & Class C& 0.46 & -2.07 & -3.39 & -2.05 & -3.19 & -4.78 & -2.89 & -3.75 & -5.39 & -3.27 & -4.36 & -6.16 & -4.54 & -5.12 & -7.14 & -6.86 & -7.86 & -10.20 & -7.46 & -8.87 & -11.50\\
  \cline{2-23}
  & Class D& 1.18 & -1.55 & -2.25 & -1.89 & -2.54 & -3.24 & -3.04 & -2.97 & -3.92 & -3.45 & -3.38 & -4.35 & -4.77 & -3.93 & -5.18 & -7.31 & -6.37 & -7.98 & -7.67 & -6.85 & -9.14\\
 \cline{2-23}
  & Class E& -5.57 & -10.52 & -11.38 & -7.78 & -12.36 & -13.03 & -7.80 & -12.30 & -13.06 & -9.05 & -13.50 & -14.04 & -10.22 & -14.08 & -14.47 & -13.04 & -16.80 & -16.73 & -13.52 & -17.56 & -17.29\\
 \cline{2-23}
  & \textbf{Average}& -2.04 & -3.20 & -3.96 & -4.11 & -4.28 & -5.24 & -4.66 & -4.76 & -5.60 & -5.17 & -5.33 & -6.31 & -6.16 & -5.88 & -7.02 & -8.23 & -8.48 & -9.79 & \textbf{-8.99} & \textbf{-9.20} & \textbf{-10.77}\\
 \hline
    \hline
  \multirow{6}{*}{RA} & Class A& -4.64 & -0.42 & -0.03 & -5.60 & -1.06 & -0.68 & -5.83 & -1.68 & -1.28 & -6.15 & -1.98 & -1.64 & -6.63 & -2.49 & -2.31 & -7.36 & -5.13 & -4.81 & -8.89 & -5.39 & -5.05\\
 \cline{2-23}
  & Class B& -2.11 & -1.39 & -2.33 & -3.69 & -2.29 & -3.82 & -4.11 & -3.28 & -4.00 & -4.54 & -3.66 & -4.90 & -5.35 & -4.04 & -5.47 & -7.53 & -7.81 & -9.88 & -8.29 & -8.23 & -10.57\\
  \cline{2-23}
  & Class C& 0.63 & -1.48 & -2.65 & -1.77 & -2.59 & -4.03 & -2.61 & -3.27 & -4.72 & -2.97 & -3.87 & -5.45 & -4.19 & -4.67 & -6.47 & -6.48 & -7.84 & -10.14 & -7.11 & -8.72 & -11.33\\
  \cline{2-23}
  & Class D& 1.73 & -0.74 & -1.43 & -1.30 & -1.56 & -2.33 & -2.59 & -2.22 & -3.23 & -2.88 & -2.65 & -3.58 & -4.24 & -3.20 & -4.50 & -6.95 & -6.04 & -7.88 & -7.26 & -7.32 & -7.25\\
   \cline{2-23}
  & Class E& -4.81 & -9.48 & -10.19 & -6.85 & -11.01 & -11.46 & -7.03 & -11.19 & -11.84 & -8.15 & -12.21 & -12.67 & -9.18 & -12.52 & -12.80 & -12.54 & -16.01 & -15.75 & -12.29 & -16.05 & -15.79\\
   \cline{2-23}
  & \textbf{Average}& -1.71 & -2.30 & -2.93 & -3.68 & -3.27 & -4.08 & -4.29 & -3.93 & -4.62 & -4.76 & -4.45 & -5.26 & -5.73 & -4.96 & -5.94 & -7.92 & -8.16 & -9.40 & \textbf{-8.57} & \textbf{-8.75} & \textbf{-9.73}\\
 \hline
     \hline
  \multirow{6}{*}{AI} & Class A& -3.61 & -1.69 & -1.74 & -4.24 & -1.99 & -2.07 & -4.43 & -2.33 & -2.31 & -4.58 & -2.19 & -2.20 & -5.01 & -2.54 & -2.69 & -6.41 & -3.74 & -3.55 & -7.27 & -3.73 & -3.79\\
  \cline{2-23}
  & Class B& -1.35 & -2.33 & -3.80 & -2.53 & -2.67 & -4.71 & -2.70 & -3.27 & -4.86 & -3.01 & -3.31 & -5.19 & -3.47 & -3.56 & -5.62 & -5.87 & -5.74 & -8.32 & -5.91 & -5.21 & -8.21\\
  \cline{2-23}
  & Class C& 1.00 & -2.48 & -3.80 & -1.49 & -3.24 & -4.96 & -2.14 & -3.85 & -5.58 & -2.49 & -3.93 & -5.81 & -3.38 & -4.60 & -6.72 & -6.11 & -6.85 & -9.39 & -5.92 & -7.41 & -10.02\\
  \cline{2-23}
  & Class D& 1.34 & -2.46 & -3.55 & -1.63 & -3.05 & -4.34 & -2.50 & -3.67 & -5.13 & -2.88 & -3.62 & -5.03 & -3.79 & -4.21 & -5.95 & -6.35 & -6.11 & -8.32 & -6.05 & -6.06 & -6.08\\
   \cline{2-23}
  & Class E& -5.13 & -8.43 & -9.08 & -6.95 & -9.79 & -10.16 & -6.95 & -9.66 & -10.19 & -7.73 & -9.80 & -10.10 & -8.79 & -10.65 & -10.78 & -11.68 & -12.88 & -12.73 & -12.29 & -13.39 & -12.74\\
   \cline{2-23}
 & \textbf{Average}& -1.36 & -3.17 & -4.13 & -3.15 & -3.79 & -4.97 & -3.53 & -4.24 & -5.35 & -3.90 & -4.25 & -5.42 & -4.62 & -4.76 & -6.09 & -6.99 & -6.71 & \textbf{-8.24} & \textbf{-7.17} & \textbf{-6.75} & -7.94\\
\hline
\end{tabular}
\end{table*}

\subsubsection{Comparison of various CNN architectures for the \hxy{adaptive-switching neural network}}
  Tables \ref{tab:difCNN} and \ref{tab:RD-difCNN} compare the Y-channel PSNR improvement and BD-rate saving of adaptive switching network with different CNN architectures over HM-16.0 baseline. 
  The following \hxy{adaptive-switching neural networks} are compared: (1) ASN@4\textit{S}, which means using the shallow model \textit{S}~\cite{vrcnn} for all CNN architectures in the adaptive switching network; (2) ASN@4\textit{D}, which is similar to ASN@4\textit{S} but using the deep model \textit{D}; (3) ASN@4$\textit{S}^{\star}$, \js{where} each CNN \js{employs} the partition-aware shallow model $\textit{S}^{\star}$; (4) ASN@4$\textit{D}^{\star}$, \js{where} each CNN uses the partition-aware deep model $\textit{D}^{\star}$ and; (5) ASN@$\textit{D}^{\star}+3\textit{S}^{\star}$, which represents a \js{hybrid combination of} the global CNN based on the deep model $\textit{D}^{\star}$ while all local CNNs use the shallow model $\textit{S}^{\star}$~\cite{vrcnn}.

We discuss the performance of the compared methods based on PSNR improvement and rate-distortion:
\begin{itemize}

\item \textbf{\js{PSNR improvement performance.}} From Table \ref{tab:difCNN} we can observe that:
(1) Our full method, ASN@4$\textit{D}^{\star}$ achieves the highest PSNR improvement (0.47 dB) over HM-16.0 baseline at QP=37. When compared with its corresponding single-CNN-based method, $\textit{D}^{\star}$ i.e., \emph{2-in+MM+AF} from Table \ref{table1}, it obtains a further 0.07 dB PSNR improvement.
\js{Results for QP=27 are of similar nature}; (2) Comparing between the results of ASN@4\textit{S} and ASN@4$\textit{S}^{\star}$, we find that the \js{partition-aware (latter)} method outperforms the \js{former} method 
by 0.08 dB at QP=37 and 0.06 dB at QP=27. \js{Interestingly}, the ASN@4$\textit{D}^{\star}$ also \js{marginally} outperforms ASN@4\textit{D}; (3) The hybrid combination method ASN@$\textit{D}^{\star}$+3$\textit{S}^{\star}$ can also obtain satisfactory results for QP=27, higher than all other methods except for ASN@4$\textit{D}^{\star}$. 


\item \textbf{Rate-distortion performance.} As mentioned in \ref{sec:test}, the flags indicating the indices of CNNs for each patch are written into the bitstream. Thus, it is necessary to evaluate the rate-distortion performance of the proposed methods. To this end, we further compare these methods in terms of BD-rate saving over HM-16.0 baseline. As shown in \ref{tab:RD-difCNN}, \js{we obtained some meaningful findings}: (1) All \hxy{adaptive-switching neural networks} outperform single-CNN-based methods - each of them can achieve higher BD-rate saving than its corresponding single-CNN-based method. \js{This} demonstrates the effectiveness of our adaptive-switching scheme. \js{Interestingly,} the gap between ASN@4$\textit{D}^{\star}$ and $\textit{D}^{\star}$ is smaller than that between ASN@4$\textit{S}^{\star}$ and $\textit{S}^{\star}$. \js{There is less room for further performance} improvement 
\js{since} the performance of $\textit{D}^{\star}$ is \js{already} quite high (9.76\%); (2) ASN@4\textit{S} can obtain a 6.94\% BD-rate saving over HM-16.0 baseline. \js{Since} the CNNs used in ASN are deeper and more complex, the proposed ASN@4\textit{D} achieves 10.49\% BD-rate saving that is clearly better than ASN@4\textit{S}; (3) 
ASN@4$\textit{S}^{\star}$ integrates our partition-aware-based approach and outperforms the original approach (ASN@4\textit{S}) by 1.54\% BD-rate savings. \js{The similar trend} is found when comparing between the \js{two deep model based methods}. 
(4) \js{The hybrid combination method}
achieves better results than their homogeneous counterparts -- ASN@4\textit{S}, ASN@4$\textit{S}^{\star}$ \& ASN@4\textit{D}. Since its time complexity on the decoder side is lower than that of ASN@4$\textit{D}^{\star}$, the ASN@$\textit{D}^{\star}$+3$\textit{S}^{\star}$ \js{presents} a practical version that is faster \js{but still obtains} satisfactory results.


\end{itemize}

Moreover, Fig.~\ref{fig:usage_rate} further shows the usage rate of our \hxy{adaptive-switching neural networks} with different CNN architectures. 

\begin{figure*}[t]
\centering
\includegraphics[height=13.35cm]{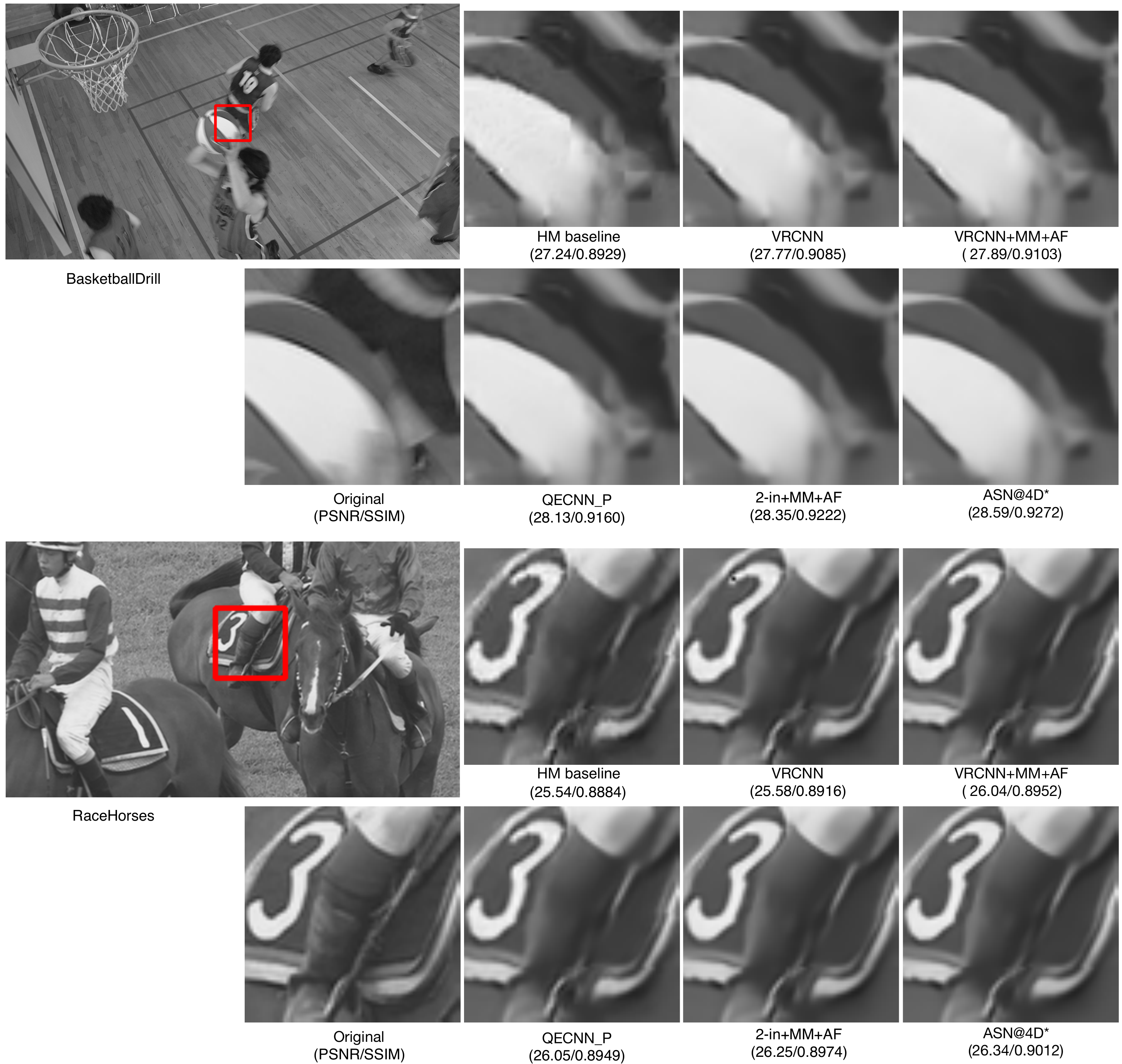}
\caption{Subjective evaluation. The decoded frames of HEVC, post-processed by CNNs with different mask generation and mask-frame fusion strategies.}
\label{fig:subjective_eval}
\end{figure*}

\subsection{Comparison \js{against} existing methods}
\subsubsection{\hxy{Rate-distortion performance}}
 \hxy{We compare our proposed methods \js{against} state-of-the-art methods in terms of BD-rate saving over HM-16.0 baseline (see Table \ref{table2} and Table \ref{table3}). Specifically, in Table \ref{table2}, we re-train all methods on our dataset and compare the performances on Classes A-E under four common configurations (cf. LP, LB, RA, AI)~\cite{ctc}.} Note that we generate training dataset under LP configuration only and use it to train models for all configurations.


\js{In Table \ref{table2}, the compared methods are as follows:}
 (1) VRCNN (\textit{S})~\cite{vrcnn} which is a \js{baseline} 
 CNN-based compressed-video post-processing method; (2) QECNN-P~\cite{qecnn} which is a compressed-video post-processing method for P frames in HEVC; \hxy{(3) DRN~\cite{DRN2019}, which is another state-of-the-art compressed-video post-processing method.} (4) VRCNN+MM+AF ($\textit{S}^{\star}$), which integrates our partition-aware-based approach into the existing \js{baseline} VRCNN method; \hxy{(5) DRN+MM+AF, which integrates our partition-aware-based approach into the existing DRN method;} (6) Our 2-in+MM+AF ($\textit{D}^{\star}$), which is the full version of our partition-aware-based approach with local mean-based mask and add-based fusion; (7) Our ASN@4$\textit{D}^{\star}$, which is the adaptive-switching scheme with \js{the deep CNN model}. From the table, we can observe that:

\begin{itemize}
\item The full version of our partition-aware CNN (\emph{Our $\textit{D}^{\star}$}) achieved the best performance over all compared single-CNN-based methods. Specifically, it can obtain over 9.76\% BD-rate reduction from standard HEVC on luminance channel under LP configuration. Similar results \js{were found} on other color channels and configurations.


\item Our proposed ASN@4$\textit{D}^{\star}$ achieved the \js{highest} BD-rate saving over HM-16.0 (up to 10.97\% on luminance channel under LP configuration). It also achieved 1.2\% BD-rate savings (on luminance channel under LP configuration) over our 2-in+MM+AF, the best single-CNN-based method. This further exemplifies the effectiveness of our proposed adaptive-switching scheme \js{in optimizing the gain locally and on the basis of the patch types}.

\item When integrating our partition-aware strategy on the existing methods, the {VRCNN+MM+AF} and {DRN+MM+AF} also obtained 3\% and \hxytwo{1.4\% BD-rate improvements over the original VRCNN and DRN methods under LP configuration. Similar gain can be obtained under other configurations.} \hxy{This also indicates that our partition-aware strategy can be flexibly integrated on the existing methods and coherently obtain performance gains.}

\item For all methods in Table~\ref{table2}, the best performance is achieved under LP configuration 
since the training data is generated under this configuration. \js{Nevertheless}, the performances under other configurations are also quite good, with only 1\%-2\% loss on average. Meanwhile, Table~\ref{table2} also shows that our methods can achieve the same or better performance on color channels (U, V) even though training was done on the luminance channel (Y).




\end{itemize}

The frames post-processed by these methods are shown in Fig.~\ref{fig:subjective_eval} to \js{assess} the \js{qualitative aspect of post-processing}. \js{Specifically,} the baseline decoded frame suffers significantly from \js{the presence of} blocking artifacts. The frames obtained by VRCNN still contain a lot of blocking artifacts, and to a much lesser extent for the case of 
frames obtained by VRCNN+MM+AF and QECNN\_P. Comparatively the proposed 2-in+MM+AF and ASN@4$\textit{D}^{\star}$ have eliminated most compression artifacts. The comparison of visual quality is consistent with the objective PSNR improvement and BD-rate saving measures. \js{This is evidential of} the effectiveness of our partition-aware CNN and adaptive-switching scheme.

\begin{table*}[ht]
\centering
\setlength\tabcolsep{2.0pt}
\caption{Comparison of different methods on Y-channel BD-rate (\%) saving over HEVC baseline under different configurations. (For compared methods DCAD and MSDD, results that reported in their papers are shown)}
\label{table3}
\begin{tabular}{|c|c|c|c|c|M{13mm}|M{13mm}|M{17mm}|}
\hline
 Conf. & Seq. & DCAD~\cite{DCAD} & MSDD~\cite{MSDD} & Our $\textit{D}^{\star}$ & Our \newline ASN@4$\textit{S}^{\star}$ & Our \newline ASN@4$\textit{D}^{\star}$ & Our \newline ASN@$\textit{D}^{\star}$+3$\textit{S}^{\star}$\\
\hline
  \multirow{5}{*}{LP} & Class B& -3.4& -5.3 & -10.1 & -9.0 & -11.1 & -11.0\\
 \cline{2-8}
  & Class C& -4.0 & -4.8 & -7.4 & -5.9 & -8.3 & -7.8\\
 \cline{2-8}
  & Class D& -5.2 & -5.2 & -7.5 & -5.9& -8.3 & -7.8\\
  \cline{2-8}
  & Class E& -7.8 & -11.8 & -14.5&  -11.9 & -15.1 & -15.0\\
  \cline{2-8}
  & \textbf{Average}& -5.0 & -6.4 & -9.7 & -8.0& \textbf{-10.5} & -10.1\\
    \hline
    \hline
  \multirow{5}{*}{LB} & Class B& -4.6 & -4.8 & -7.6 & -6.5&  -8.3 & -8.0\\
 \cline{2-8}
  & Class C& -4.3 & -6.1 & -6.9 & -5.3& -7.5 & -7.0\\
 \cline{2-8}
  & Class D& -4.4  & -6.6 & -7.3 & -5.3& -7.7 & -7.9\\
  \cline{2-8}
  & Class E& -10.1  & -8.7 & -13.0 & -10.3& -13.5& -12.8\\
  \cline{2-8}
  & \textbf{Average}& -5.5  & -6.4 & -8.4 & -6.6& \textbf{-8.9} & -8.4\\
 \hline
    \hline
  \multirow{5}{*}{RA} & Class B& -4.6  & -3.8 & -7.5 & -6.5& -8.3& -8.2\\
 \cline{2-8}
  & Class C& -4.5 & -5.8 & -6.5 & -5.0& -7.1& -6.9 \\
 \cline{2-8}
  & Class D& -4.4 & -6.5 & -7.0 & -4.9& -7.3& -6.8\\
  \cline{2-8}
  & Class E& -10.1 & -9.6 & -12.5 & -9.5& -12.3& -11.7\\
  \cline{2-8}
  & \textbf{Average}& -5.5 & -6.3 & -8.1 & -6.3& \textbf{-8.5} & -8.2\\
 \hline
     \hline
  \multirow{5}{*}{AI} & Class B& -3.4 & -3.8 & -5.9 & -4.8&  -5.9 & -6.1\\
 \cline{2-8}
  & Class C& -4.6 & -5.8 & -6.1 & -4.6& -5.9& -5.9\\
 \cline{2-8}
  & Class D& -5.2 & -6.5 & -6.4 & -4.7& -6.1& -5.8\\
  \cline{2-8}
  & Class E& -7.8 & -9.6 & -11.7 & -9.5& -12.3& -12.0\\
  \cline{2-8}
  & \textbf{Average}& -5.0 & -6.5 & \textbf{-7.1} & -5.6& \textbf{-7.1} & \textbf{-7.1}\\
\hline
\end{tabular}
\end{table*}

\begin{figure}[t]
 \centering
 \includegraphics[width=0.9\linewidth]{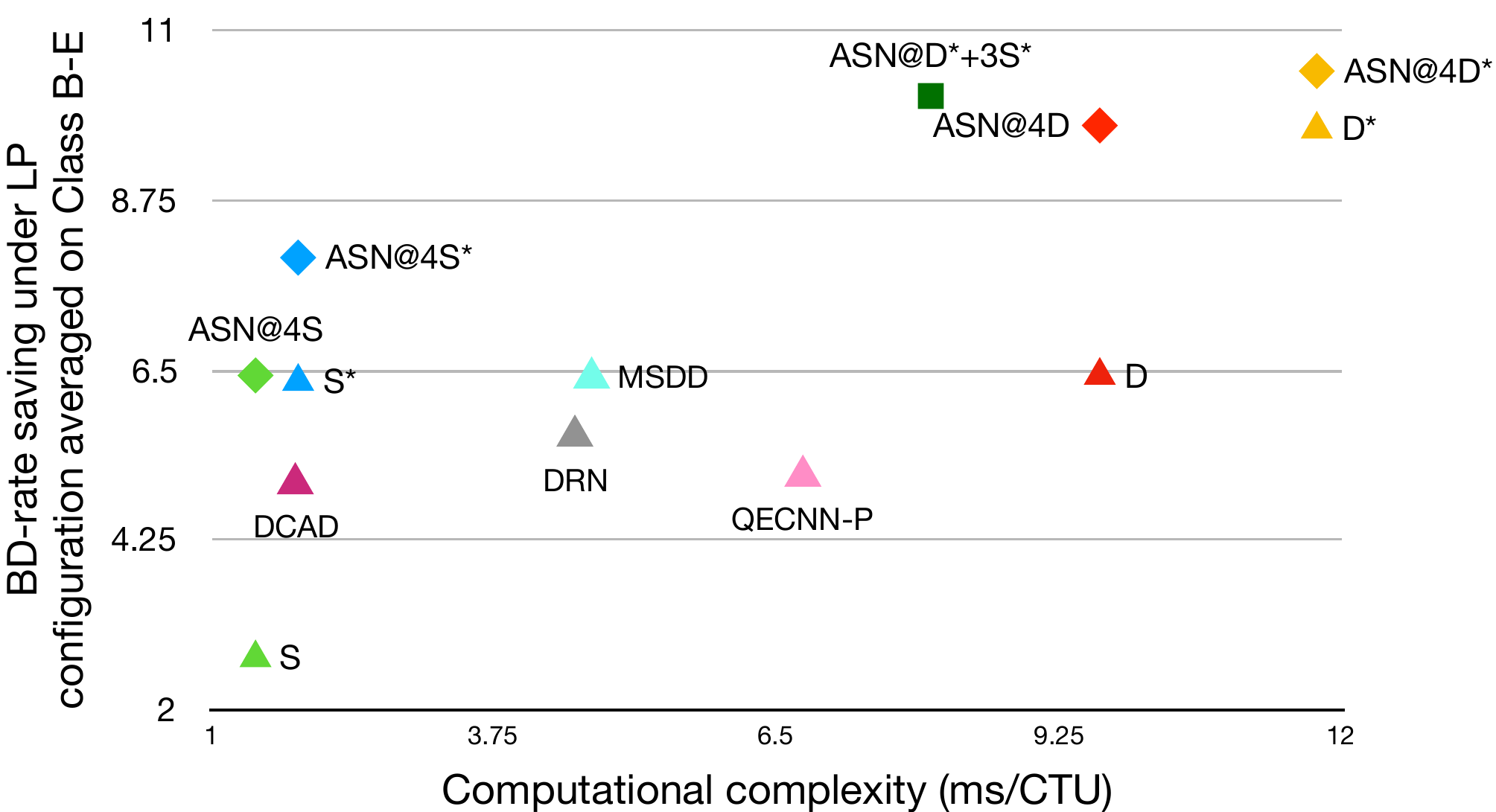}
 \caption{\hxy{Comparison of different methods on computational time per CTU in decoder side versus BD-rate saving over HEVC baseline.}}
 \label{fig:timevsBD}
\end{figure}

\hxy{In order to further evaluate our approach,
Table \ref{table3} further shows the performance comparison between our approach and more state-of-the-art compressed-video post-processing methods~\cite{DCAD, MSDD}. In Table \ref{table3}, since the source codes of the compared methods~\cite{DCAD, MSDD} are not available, we list the BD-rate savings reported in their papers and compare them with our approach. }

\hxy{Here, four versions of our approach are listed: (1) Our 2-in+MM+AF ($\textit{D}^{\star}$), which is the full version of our partition-aware-based approach with local mean-based mask and add-based fusion; (2) ASN@4$\textit{S}^{\star}$, which is the adaptive-switching scheme with each CNN employs the partition-aware shallow model $\textit{S}^{\star}$; (3) ASN@4$\textit{D}^{\star}$, where each CNN uses the partition-aware deep model $\textit{D}^{\star}$ and; (4) ASN@$\textit{D}^{\star}+3\textit{S}^{\star}$, which represents a hybrid combination of the global CNN based on the deep model $\textit{D}^{\star}$ while all local CNNs use the shallow model $\textit{S}^{\star}$.}

\hxy{From Table \ref{table3}, we can observe that our proposed $\textit{D}^{\star}$, ASN@4$\textit{D}^{\star}$, and ASN@$\textit{D}^{\star}+3\textit{S}^{\star}$ methods outperform the existing methods~\cite{DCAD, MSDD} under all configurations. Besides, by integrating our partition-aware and adaptive-switching strategies on a shallow model~\cite{vrcnn}, our ASN@4$\textit{S}^{\star}$ method can also obtain satisfactory performances, which has better results than {DCAD} and 
{MSDD} under LP configuration and similar results under other configurations. This further demonstrates the effectiveness of our approach.} Note that our approaches show higher improvements under LP configuration since the training dataset is generated under this configuration. In practice, more improvements can also be obtained for other configurations if generating training data for each configuration respectively.


\subsubsection{Complexity analysis in decoder side}
In this part, we evaluate the computational complexity of our proposed methods and the existing techniques. Following the other post-processing methods~\cite{qecnn}, we evaluate the computational complexity by the running time per Coding Tree Unit (CTU) in the decoder side. Experiments \js{were} conducted using one GeForce GTX 1080Ti GPU. Fig. \ref{fig:timevsBD} shows the running time per CTU for different methods versus BD-rate saving averaged on Class B-E under LP configuration. We observe from Fig. \ref{fig:timevsBD} that:
\begin{itemize}

\item All our \hxy{adaptive-switching neural networks} achieved better BD-rate savings than their single-CNN-based counterparts without costing extra computational time in decoder side.

\item \hxy{\js{As expected, the incorporation of partition awareness into our CNNs comes at the expense of higher computational cost.} The $\textit{D}^{\star}$ achieved 2.84\% BD-rate saving over \textit{D} at the expense of extra 2.11 ms per CTU. \js{But for the case of shallow models,} $\textit{S}^{\star}$ achieved a 3\% BD-rate saving over \textit{S} at the expense of only extra 0.41 ms per CTU. }

\item Our proposed ASN@4$\textit{D}^{\star}$ achieved up to 10.74\% BD-rate saving. However, it is also the slowest one, taking 11.8 ms when processing one CTU. The fastest method is our ASN@4\textit{S} which achieved 1.43 ms per CTU. Its running time is similar to that of \textit{S}, but it achieved a 3.13\% BD-rate saving over VRCNN. \hxy{Moreover, ASN@4$\textit{S}^{\star}$ has lesser complexity than most of the compared methods (QECNN-P, DRN, MSDD) while obtaining better
performances.}

\item \js{For practical considerations, the hybrid} ASN@$\textit{D}^{\star}$+3$\textit{S}^{\star}$ approach is only marginally worser than ASN@4$\textit{D}^{\star}$ in terms of BD-rate savings but faster \js{by a larger extent}. In addition, it outperforms ASN@4\textit{D} and also is faster. \hxy{When compared with the existing methods, ASN@$\textit{D}^{\star}$+3$\textit{S}^{\star}$ outperforms VRCNN, QECNN-P, DCAD, DRN, MSDD by a large margin with slightly higher complexity than QECNN-P.} 
Note that the performance of ASN@$\textit{D}^{\star}$+3$\textit{S}^{\star}$ in decoder is related to the test data.
As some patches are not \js{flagged} to use $\textit{D}^{\star}$ but $\textit{S}^{\star}$ instead, therefore speedup \js{can be} achieved.
\end{itemize}

Note that the batch size is only set to 1 in our experiments. The average computational complexity per CTU can be further reduced by utilizing larger batch sizes together with some parallel processing architectures~\cite{zhang2012efficient,wang2018collaborative}.

\section{Conclusion}
\js{In this work, \hxy{we present a number of techniques to address neural network based post-processing for HEVC}, reporting significant findings and improvements.} Our partition-aware network utilizes the partition information that has already \js{existed} in the bitstream to design a mask and integrate it with the decoded frame \js{as input} to the CNN. 
\js{To guide the post-processing,}
we also propose an adaptive-switching scheme which consists of multiple carefully trained CNNs, \js{which are aimed to} adaptively handle variations in content and distortion within compressed-video frames. Experimental results show that our partition-aware CNN is more effective compared to other single-CNN-based methods, and our adaptive-switching scheme \js{is robust in bringing} further improvement to our proposal. \js{We also made publicly a large-scale dataset to facilitate future research in this direction.}
\label{sec:conclu}




\bibliographystyle{IEEEtran}
\bibliography{IEEEabrv,refs.bib}

\end{document}